\documentclass[10pt,pra,aps,twocolumn,superscriptaddress,floatfix,showpacs]{revtex4-2}
\usepackage[T1]{fontenc}
\usepackage{graphicx,amsfonts,amssymb,amsmath,mathrsfs,bm,dsfont}
\usepackage{braket}
\usepackage{anyfontsize}
\usepackage{mathtools}
\usepackage{amsthm}
\usepackage{physics}
\usepackage{stmaryrd}
\usepackage{appendix}
\usepackage{minitoc}

\usepackage{multirow}
\usepackage{hhline}
\usepackage{hyperref}
\usepackage{ulem}
\usepackage[ruled]{algorithm2e}
\hypersetup{ 
    colorlinks=true,
    linkcolor=blue,
    breaklinks=true,
    citecolor=blue,
    filecolor=blue,
    urlcolor=blue}
\usepackage{url}

\usepackage{comment}
\usepackage[colorinlistoftodos]{todonotes}

\newcommand{\Graph}{{\mathcal{G}}}

\newcommand{\vertices}{{\mathcal{V}}}

\definecolor{darkgreen}{RGB}{15, 30, 35}
\definecolor{softgreen}{RGB}{23, 48, 53}
\definecolor{brightgreen}{RGB}{225, 246, 233}
\definecolor{mintgreen}{RGB}{0, 200, 135}
\definecolor{metalblue}{RGB}{57, 115, 120}
\definecolor{neonblue}{RGB}{146, 200, 229}
\definecolor{pasqalpurple}{RGB}{134, 123, 250}
\definecolor{pasqalorange}{RGB}{255, 152, 110}

\date{\today}

\begin{document}

\title{Identifying hard native instances for the maximum independent set problem on neutral atoms quantum processors}

\author{Pierre Cazals}
\email{pierre.cazals@pasqal.com}
\author{Aymeric François}
\author{Loïc Henriet}
\author{Lucas Leclerc}
\affiliation{Pasqal, 24 rue Emile Baudot, 91120 Palaiseau, France}
\author{Malory Marin}
\affiliation{ENS de Lyon, CNRS, Université Claude Bernard Lyon 1, LIP, UMR 5668, 69342 Lyon Cedex 07, France}
\author{\\Yassine Naghmouchi}
\author{Wesley da Silva Coelho}
\affiliation{Pasqal, 24 rue Emile Baudot, 91120 Palaiseau, France}
\author{Florian Sikora}
\affiliation{LAMSADE, Université Paris-Dauphine, PSL University, CNRS, 75016 Paris, France}
\author{Vittorio Vitale}
\affiliation{Pasqal, 24 rue Emile Baudot, 91120 Palaiseau, France}
\author{\\Rémi Watrigant}
\affiliation{ENS de Lyon, CNRS, Université Claude Bernard Lyon 1, LIP, UMR 5668, 69342 Lyon Cedex 07, France}
\author{Monique Witt Garzillo}
\author{Constantin Dalyac}
\email{constantin@pasqal.com}
\affiliation{Pasqal, 24 rue Emile Baudot, 91120 Palaiseau, France}

\begin{abstract}
The Maximum Independent Set (MIS) problem is a fundamental combinatorial optimization task that can be naturally mapped onto the Ising Hamiltonian of neutral atom quantum processors. Given its connection to NP-hard problems and real-world applications, there has been significant experimental interest in exploring quantum advantage for MIS. Pioneering experiments on King’s Lattice graphs suggested a quadratic speed-up over simulated annealing, but recent benchmarks using state-of-the-art methods found no clear advantage, likely due to the structured nature of the tested instances. In this work, we generate hard instances of unit-disk graphs by leveraging complexity theory results and varying key hardness parameters such as density and treewidth. For a fixed graph size, we show that increasing these parameters can lead to prohibitive classical runtime increases of several orders of magnitude. We then compare classical and quantum approaches on small instances and find that, at this scale, quantum solutions are slower than classical ones for finding exact solutions. Based on extended classical benchmarks at larger problem sizes, we estimate that scaling up to a thousand atoms with a 1 kHz repetition rate is a necessary step toward demonstrating a computational advantage with quantum methods.
\end{abstract}

\maketitle

\section*{Introduction}
Combinatorial optimization encompasses a wide range of problems whose goal is to find the best solution from a finite but often exponentially large set of possibilities.
Examples of such problems naturally appear across several domains, including logistics, telecommunications, finance, or healthcare~\cite{paschos2014applications,resende2003combinatorial,du2019real,ali2015mathematical}. 
Given their computational complexity, classical algorithms often struggle to solve large instances efficiently, which has led to interest in exploring whether quantum computing could provide a speed-up for such tasks, fueled initially by results in complexity theory like Shor’s algorithm~\cite{Shor1995PolynomialTimeAF} or the supremacy claim of the Quantum Approximate Optimization Algorithm (QAOA)~\cite{farhi2014quantum}.
Complexity theory, which focuses on distinguishing problems that can be solved efficiently from those that cannot, does indeed provide valuable insight into the potential for quantum advantage in optimization, but formal complexity-theoretic separations are neither necessary nor sufficient for demonstrating a practical speed up~\cite{abbas2023quantum}. A broader perspective is needed where one considers realistic problem instances where quantum algorithms might still offer exponential improvements. Recent empirical studies suggest for example that carefully designed problem instances, tailored to specific quantum architectures, could enable significant acceleration in combinatorial optimization tasks~\cite{tasseff2024emerging}.

A particular class of combinatorial optimization problem that is native to neutral atoms is the Maximum Independent Set problem on unit-disk graphs (UD-MIS)~\cite{ebadi2021quantum, dalyac2024graph, dalyac2021qualifying, byun2022finding, leclerc2024implementingtransferableannealingprotocols}.
It involves determining the largest subset of vertices in a graph such that neither vertex is connected by an edge.
Remarkably, it can be translated into an optimization problem encoded in the ground-state of an Ising-like Hamiltonian~\cite{lucas2014ising, pichler2018quantum} and is therefore native to neutral atoms quantum platforms. In practice, by positioning atoms at specified coordinates, their spatial arrangement naturally enforces the independence constraint in UD graphs, i.e. two nearby atoms cannot simultaneously be excited to Rydberg states~\cite{lukin2001dipole, browaeys2020many}. This mechanics, dubbed Rydberg blockade~\cite{saffman2010quantum, urban2009observation,labuhn2016tunable}, allows one to realize in practice unit-disk (UD) graphs~\footnote{A unit-disk graph is defined by the positions of nodes, where an edge exists between two nodes if the distance between them is less than a constant \( r \), typically set to \( r = 1 \).}.
To embed more generic classes of graph, encoding techniques employing Rydberg wires have been proposed ~\cite{kim2022rydberg}, e.g. in the case of platonic graphs~\cite{byun2022quantum} or bounded-degree graphs~\cite{dalyac2023exploring}. 
Further approaches aim to encode graphs with arbitrary connectivity using reductions that map them onto natively embeddable UD-graphs~\cite{nguyen2023quantum, lanthaler2023rydberg, byun2024rydberg, jeong2023quantum}. 
Eventually, once the embedding is performed, the remaining task consists in preparing the quantum state that encodes the solution of the optimization problem by leveraging the quantum dynamics and extracting the solution through projective measurements.

Early investigations have shown that variational quantum annealing~\cite{finnila1994quantum, kadowaki1998quantum,farhi2000quantum,santoro2002theory} outperforms QAOA~\cite{QAOA,Blekos_2024,Zhou20} in finding the ground-state that encodes the solution to the MIS~\cite{ebadi2022quantum}. Further works even identified a quadratic speed-up of quantum annealing with respect to classical annealing~\cite{schiffer2024circumventing} on King's Lattice (KL) graphs. However, careful benchmarks have provided no clear speed-up with respect to other state-of-the-art classical methods~\cite{andrist2023hardness}. Historically, similar benchmarks had tamed the hopes of the quantum annealing community, based on tests on first-generation devices~\cite{selby2014efficient,mandra2016strengths,albash2018demonstration}. However, recent studies have shown encouraging results thanks to the latest hardware improvements and the identification of hard native instances for superconducting platforms~\cite{tasseff2024emerging}. 
 
So far, any sign of a quantum speed-up or advantage in combinatorial optimization with neutral atoms remains quite elusive. Primarily, we believe that it is because the instances studied have a very simple structure, allowing for both traditional resources and quantum computers to solve them with minimal efforts~\cite{ronnow2014defining, dalzell2020many}; secondly, due to the limitations imposed by the small number of qubits of current devices.
For these reasons, it is imperative to design \textit{native} benchmarks that employ the maximum number of available atoms and that are, at the same time, as hard as possible for classical algorithms~\cite{katzgraber2015seeking,hen2015probing,zhu2016best}.
Therefore, the goal of the present work is to determine instances embeddable on neutral atoms QPUs that pose significant challenges for classical solvers. If so, we aim to identify the hardest instances that align with today's capabilities, thereby raising the bar for classical solvers and comparing them with their quantum counterpart.  

To this end, we proceed as follows:
In Sec.~\ref{sec:summary}, we draw a summary of the previous contributions to the literature of quantum optimization and describe our main findings.
In Sec.~\ref{sec:complexity_theory}, we provide a more detailed analytical understanding for the hardness of solving UD-MIS in a classical setting, and we identify explicit hardness parameters.
In Sec.~\ref{sec:numerics}, we test our theoretical predictions employing numerical simulations using a classical branch-and-bound solver~\cite{cplex2024}. 
Finally, in Sec.~\ref{sec:experiments}, we provide numerical and experimental results employing a neutral atom quantum platform to investigate current capabilities and trace the path for future hardware and algorithmic developments.
We also provide more details in the appendices on how to map UD-MIS to neutral atoms, the annealing procedure, and the error mitigation post-processing we employ.

\begin{figure*}
    \centering
    \includegraphics[width=\linewidth]{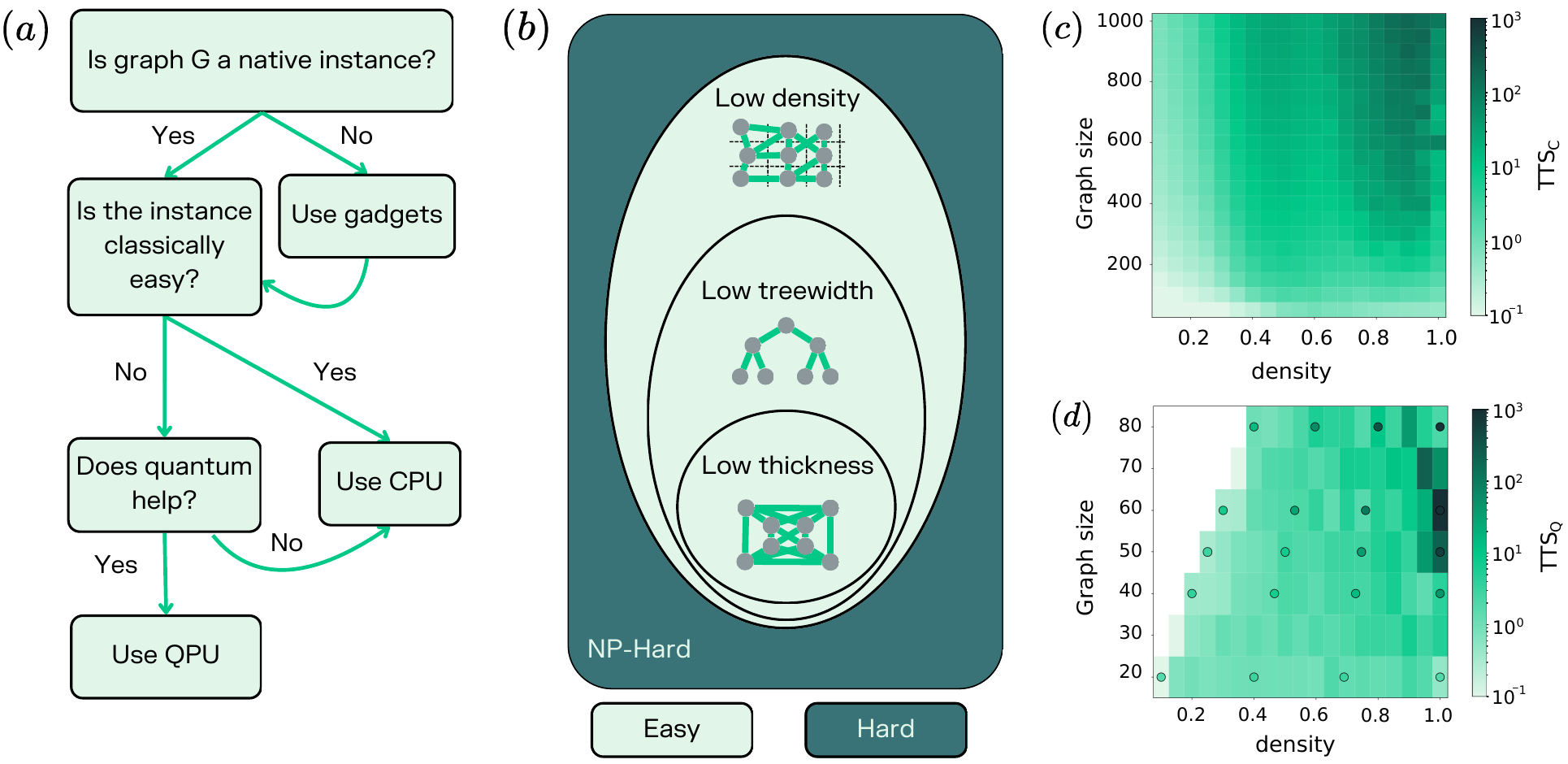}
    \caption{Overview of the paper. We investigate the hardness of the MIS problem on natively embeddable graphs. 
    $(a)$ According to our reasoning, we focus on UD graphs, since providing evidence of quantum speed up on these is essential for tackling problems on graphs with arbitrary connectivity. $(b)$ We provide analytical understanding that UD graphs with low treewidth, density or thickness, represent classes of graphs that are classically easy to solve (Sec.~\ref{sec:complexity_theory}). The three parameters are hierarchically nested, as explained in Sec. \ref{sec:param}. $(c)$-$(d)$ We show numerical results for the MIS problem on graphs that are natively embeddable on the hardware. For details on the generation of the graphs, with increasing densities, the reader may refer to App.~\ref{sec:generate_instances}.
    In $(c)$, we provide classical simulations showing the time to solution $\mathrm{TTS}_C$ for solving the MIS averaged over $50$ random graphs for each value of size and density. TTS$_C$ is measured in CPLEX deterministic ticks~\cite{cplex2024}.
    In $(d)$ we represent the time to solution according to numerical simulations of the quantum annealing procedure implemented TTS$_Q$, while the colored circles depict experimental results obtained with an Orion Alpha class neutral atom quantum processor from Pasqal. The numerical (experimental) results are averaged over $20$ ($10$) instances with $1000$ ($500$) sampled bitstrings each.}
    \label{fig:fig1}
\end{figure*}

\section{Overview of the work}
\label{sec:summary}
In this section, we provide an overview of our work.
In Sec.~\ref{subsec:pb_form}, we define the weighted version of the MIS problem and the metrics employed to evaluate the performances of our solvers. In Sec.~\ref{sec:prior_works}, we discuss previous works aiming at analyzing the hardness of the optimization problem.
Finally, in Sec.~\ref{sec:our_contr} we outline our contribution, aided by Fig.~\ref{fig:fig1}.

\subsection{Problem formulation and metrics}
\label{subsec:pb_form}

The Maximum Weighted Independent Set (MWIS) problem is a well-known optimization problem in graph theory.
Let us consider a graph \(G = (V, E)\), where \(V\) represents the set of vertices and \(E\) is the set of edges that connect pairs of vertices. An independent set in this graph is a set of vertices such that no two vertices in the set are connected by an edge. In other words, for every pair of vertices \(i\) and \(j\), if there is an edge between them, then at least one of these vertices cannot be part of the independent set. 
In the MWIS problem, we aim to find the independent set that not only satisfies this condition but also has the maximum total weight. Each vertex \(i\) has an associated weight \( w_i \), and the goal is to select the vertices with the highest weights, subject to the constraint that no two selected vertices are adjacent.
To make this more formal, we define binary variables \( x_i \) for each vertex \( i \), where \( x_i = 1 \) if vertex \( i \) is included in the independent set, and \( x_i = 0 \) otherwise. The MWIS problem can then be formulated as an optimization problem, as detailed in Sec.~\ref{sec:prob_statement}.

Complexity theory plays a key role in assessing the difficulty of MIS and other optimization problems. For  instance, complexity theory teaches us that the difficulty of a problem can depend on the type of graph on which it is defined. While MIS is generally challenging, it admits an exact polynomial-time algorithm when restricted to tree graphs~\cite{bhattacharya2014maximum}. In contrast, MIS remains NP-hard for both Erdős–Rényi (ER) graphs~\cite{garey1979computers} and Kings Lattice (KL) graphs~\cite{ebadi2022quantum}.
Yet, the NP-hard classification does not always fully capture the practical difficulty: on average, solving MIS on KL graphs is significantly easier than on ER graphs, with state-of-the-art classical solvers efficiently handling graphs with thousands of nodes in minutes, without encountering exponential runtime growth~\cite{andrist2023hardness}. This discrepancy highlights the limitations of the worst-case complexity analysis, which may not always align with the empirical difficulty of solving problem instances.
For the reasons above, we introduce here some figures of merit that are useful to evaluate the performances of classical and the quantum algorithms when employed in practice.
In the case of classical solvers, we consider the time-to-solution, that we denote TTS$_C$. The latter gives an estimate of the
time needed to find the optimal solution 
of the MIS problem. In the following, it will be measured in terms of number of steps (or memory access) made by the solver during the optimization process.
In the quantum case, we employ two quantities: a suitable definition of the time-to-solution, and the gap to solution.
The former, TTS$_Q$, is equivalent to the number of shots required to find the exact
solution with 99\% success probability~\cite{andrist2023hardness}:
\begin{equation}
\label{eq:tts}
    \text{TTS}_Q=\frac{\log(1-0.99)}{\log(1-P_{\mathrm{MIS}})},
\end{equation}
where $P_{\mathrm{MIS}}$ is the probability of finding an MIS.
The latter corresponds to the normalized distance between the optimal solution and the obtained one 
\begin{equation}
    \Delta(x)=\left| \frac{x-x^{*}}{x^{*}}\right|.
\end{equation}
In the following, we will employ these performance metrics for supporting and benchmarking our theoretical insights, coming from complexity theory, with the goal to define hard native instances for the MIS problem. Before this, we will summarize previous contributions to the topic in the following section.

\subsection{Prior works}
\label{sec:prior_works}
Understanding the complexity of combinatorial optimization problems in the most diverse settings has been a hot topic in the past decades.
From a physics perspective, the topic has been tackled, drawing parallels between spin glasses and constraint satisfaction problems. The latter has revealed phase transitions in problem complexity, where control parameters influence the transition from easy to hard problems, as observed in phenomena like the clause-to-variable ratio in SAT instances~\cite{mezard1986replica, fu1986application, monasson1999determining}. These transitions align with structural changes in the solution space, such as the fragmentation of solution clusters~\cite{krzakala2007gibbs,achlioptas2008algorithmic,coja2015independent}. Instances generated close to these transitions are typically hard instances~\cite{hartmann2006phase, gent1994sat}.

In computer science, hard instances are often obtained by employing either a generate-and-test method~\cite{culberson2001frozen,mammen1997new,vlasie1995systematic,vlasie1996very} or constructive approaches, with the particular MIS case having been extensively studied~\cite{bang2019haj, leo2013solving,sloane_graphs, dimacs1992}.

In the quantum computing community, the commonly used approach has been cherry-picking hard instances with respect to the native constraints imposed by the quantum architecture, either by generate-and-test methods\,\cite{marshall2016practical}, using constructive methods such as planted solutions\,\cite{somma2012quantum}, or corrupted biased ferromagnets\,\cite{pang2021potential,tasseff2024emerging} in the case of superconducting adiabatic quantum
optimization system~\cite{berkley2010scalable}.

For neutral atoms, recent works have focused on benchmarking quantum annealing against classical counterparts.
In Ref.~\cite{ebadi2022quantum}, a benchmark of quantum variational algorithms against simulated annealing to solve the MIS on UD-graphs has been performed, showing a limited quantum speed-up. 
To extend the comparison to a larger family of classical algorithms, the authors in Ref.~\cite{andrist2023hardness} complemented the previous results by employing both exact and heuristic classical algorithms, showing that classical solvers show better performances with respect to the quantum algorithm.
Finally, while one might attempt to use polynomial-time reductions to encode problems on arbitrary graphs—such as embedding ER graphs into a neutral atom platform via gadgets~\cite{nguyen2023quantum}—such reductions transform the problem into an UD instance. As a result, studying native instances directly is more informative than relying on indirect encoding (Fig.~\ref{fig:fig1}$(a)$).

\subsection{Our contribution}
\label{sec:our_contr}
Our contribution to the topic is highlighted in Fig.~\ref{fig:fig1}. 
In this work, we aim to extend previous results by incorporating theoretical insights from complexity theory. We examine strategies developed by complexity theorists to address the challenges posed by NP-hardness~\cite{garey1979computers} and we show that these approaches can, in some cases, uncover the significance of intrinsic parameters, affecting the computational difficulty of these problems. Specifically, we focus on identifying potential hardness parameters, which often reflect structural properties, that govern the complexity of solving optimization problems. While the existence of distinct hardness regimes cannot always be guaranteed, our approach allows us to delineate the conditions under which problem instances are computationally tractable. Through this analysis, we uncover the factors that influence complexity and provide guidelines for generating harder UD-MIS instances.
As shown in Fig.~\ref{fig:fig1}$(b)$, we identify three parameters that can be used to tune the hardness of optimization problem on UD-graphs: density, thickness~\cite{van2005approximation} and treewidth\,\cite{arnborg1989linear}. 
We observe that among NP-hard problems, which are in general computationally hard, the ones characterized by a small value of the above-mentioned parameters are the ones that are classically the easiest to solve. 

In Fig.~\ref{fig:fig1}$(c)-(d)$, we show, as an example, numerical simulations for the case of the MIS problem on graphs of increasing density. The graphs are generated by sampling the nodes of a triangular lattice, as shown in App.~\ref{sec:generate_instances}. Increasing density means increasing the number of nodes sampled over the total number of possible sites considered. 
In Fig.~\ref{fig:fig1}$(c)$ we show the results obtained by employing a branch-and-bound solver (CPLEX,~\cite{cplex2024}), as it is discussed in Sec.~\ref{sec:branchandbound}. We plot a color map of the classical time-to-solution (TTS$_C$) as a function of the size $N$ and the density of the graph.
TTS$_C$ is measured in number of operations (ticks), and hence is independent of the hardware performances.
We observe that larger and more dense graphs present a harder challenge to the solver, as expected. 
In Fig.~\ref{fig:fig1}$(d)$ we show an equivalent plot employing a quantum solver. We plot the quantum time-to-solution TTS$_Q$ as defined before. 
Employing this quantity, both Fig.~\ref{fig:fig1}$(c)$ and $(d)$ are independent of the actual speed of the processor. 
Geometric limitations prevent the embedding on the neutral atom hardware of the low-density easiest instances.  

We observe that, as density and system size increase, the instances become progressively more difficult, exhibiting a qualitative behavior similar to that shown in Fig.~\ref{fig:fig1}$(c)$.
The experimental data demonstrates performance comparable to the simulated one for small enough systems, but struggles to keep up with both density and system size increasing. All the experimental results are shown after applying an error mitigation and post-processing procedure that is described in App.~\ref{sec:nm_and_pp}.

\section{Identification of hard instances}
\label{sec:complexity_theory}

The goal of this section is to identify the inherently hard instances for the MIS problem, keeping in mind that we are interested in investigating graphs compatible with neutral atoms quantum platforms.

\subsection{Overcoming NP-hardness}
  
An effective way to address the computational challenges of NP-hard problems is through the study of sub-exponential algorithms. Those algorithms provide running times slower than any exponential function of the input size $n$, typically of the form $2^{O(n^\alpha)}$ with $\alpha < 1$. For example, a sub-exponential algorithm for the MIS problem on planar graphs is the $2^{\mathcal{O}(\sqrt{n})}$-algorithm that leverages the planar separator theorem, i.e., it partitions the graph into smaller subgraphs using separators of size $O(\sqrt n)$~\cite{lipton1977applications}. Similar techniques apply also to UD graphs, which share structural properties with planar graphs, enabling a $2^{\mathcal{O}(\sqrt{n})}$ sub-exponential algorithm for UD-MIS~\cite{de2018framework}. This framework extends to intersection graphs of similarly-sized fat~\footnote{An object is said to be $\alpha$-fat if the ratio between its circumscribed radius and inscribed radius is at most $\alpha$. Intuitively, an $\alpha$-fat object cannot be excessively elongated or thin. A set of objects is said to be similarly-sized if the ratio between the diameters of the largest and smallest object in the set is bounded by some constant $\alpha > 0$.} objects in higher dimensions, enabling algorithms with running times of $2^{O(n^{1-\frac{1}{d}})}$ for fixed dimensions $d\geqslant 2$.
 
Although this result suggests that UD-MIS is relatively easier to solve exactly compared to other hard problems, it still entails an exponential dependence on the instance size. To deal with this, two primary strategies are often considered: shifting the exponential dependence onto other parameters, or compromising on solution quality by accepting approximate solutions. In the following, we focus on these two approaches within the context of the UD-MIS problem, and we will see that these methods can sometimes explicit an exponential dependence that is not related to the size of the problem, but to some inner parameter of the graph. Therefore, the final task shall be to identify them to understand how they affect the inner complexity of solving the UD-MIS problem.

\subsection{Parameterized complexity}
\label{sec:param}

Parameterized complexity~\cite{cygan2015parameterized} is a modern branch of complexity theory that provides a more refined way to analyze the difficulty of computational problems. Rather than evaluating the running time based solely on the input size, parameterized complexity introduces one or more parameters associated with the problem instance. The central concept in this framework is the fixed-parameter tractability (FPT), in which an algorithm is considered FPT if it runs in time 
\begin{equation}
    \mathcal{O}(f(k) \cdot n^c),
\end{equation}  
where $n$ is the input size, $k$ is the parameter, $c$ is a constant, and $f$ is any computable function, usually exponential in the parameter $k$.  

When considering the MIS problem, one powerful approach is to parameterize it by the \textit{treewidth} $w$ of the input graph~\cite{arnborg1989linear}. Informally speaking, the treewidth measures how close a graph is to being a tree. Specifically, given a tree decomposition of width $w$ of a graph of size $n$, the MIS problem can be solved in time 
\begin{equation}
    \mathcal{O}(2^{w} \cdot w^{\mathcal{O}(1)} \cdot n).
\end{equation}
Hence, although the problem is itself NP-hard, if the specific graph instance has small treewidth then the exact solving time is tractable, specifically by generalizing the polynomial-time algorithm designed for tree instances~\cite{lipton1977applications}. In Sec.~\ref{sec:numerics}, we study how the complexity of the MIS problem on UD-graphs instances scales in practice by artificially increasing the treewidth.

In the particular case of UD-graphs, another fixed-parameter result is based on the concept of thickness. The thickness of a UD graph is defined as the minimum number of disk centers (or vertices) contained in any slab within a slab decomposition, where slabs are defined as regions between parallel lines spaced one unit apart. This measure of thickness can be computed in polynomial time~\cite{th:slabDecompoPolyvan2004optimization}.
This geometric characterization of thickness directly influences the computational complexity of solving UD-MIS: in Ref.~\cite{van2005approximation} the authors propose an FPT algorithm to find the MIS of a UD graph of thickness $t$ in time
\begin{equation}
    \mathcal{O}(t^2 \cdot 2^{2t} \cdot n).
\end{equation} 
Thus, at fixed thickness, the runtime of solving UD-MIS grows only linearly with the graph size, providing useful insight into why certain instances, such as those with low thickness, are easier to solve. As with treewidth, we can focus our attention on building graphs with high thickness and showcase the performance of classical solvers.
In Ref.~\cite{van2005approximation} the authors also show that the thickness is connected to the density of a graph (defined as the maximum number of disk centers in an arbitrary unit grid). Therefore, in Sec.~\ref{sec:numerics}, we will study in particular how the hardness behaves as a function of the density.

More generally, a structural hierarchy exists between thickness, treewidth, and density in UD graphs. Bounded thickness implies bounded treewidth, as small thickness constrains the size of separators, which in turn limits treewidth. Additionally, bounded treewidth implies bounded density, since it enforces a bound on the clique number, which directly restricts the graph’s density. Consequently, graphs of bounded thickness also have bounded density, reinforcing the dependency between these parameters.

\subsection{Approximate solutions}

Another way to address the computational intractability of NP-hard problems is to compromise on solution quality. An algorithm is said to be a $\lambda$-approximation (for some real $\lambda > 1$) for a maximization problem if it runs in polynomial time and always returns a solution whose value is at least $1/\lambda$ of the optimum.  

In arbitrary graphs, designing an approximation algorithm for the MIS problem is not feasible. More formally, for any $\varepsilon > 0$, there cannot exist a $n^{1-\varepsilon}$-approximation algorithm for MIS, unless P = NP~\cite{haastad1999clique}. However, on UD graphs, simple strategies achieve non-trivial approximations. For example, the greedy algorithm that iteratively selects the leftmost disk removes its neighborhood, continues recursively on the remaining graph, and produces an IS whose size is at least $1/3$ of the MIS of the input graph~\cite{hochbaum1985approximation}. Remarkably, this can be improved significantly. For any $\varepsilon > 0$, there exists a $(1+\varepsilon)$-approximation algorithm for UD-MIS, running in time $n^{O(1/\varepsilon)}$. Such a family of algorithms is known as a Polynomial-Time Approximation Scheme (PTAS). 
In the case of UD-MIS, it is proved~\,\cite{van2005approximation} that, for any $\varepsilon > 0$, there exists a $(1 + \varepsilon)$-approximation algorithm for UD-MIS, running in time 
\begin{equation}
    \mathcal{O}(d^{\mathcal{O}(1/\varepsilon)} \cdot n),
\end{equation}
where $d$ is its density, defined as before, i.e. the maximum number of disk centers in the unit grid, oriented to minimize this value. Note that this definition is closely related to the concept of a maximum clique, and the two are linearly correlated. Importantly, the exponential dependency on $1/\varepsilon$ is with respect to the density $d$, rather than the number of vertices $n$.

\subsection{Conclusions}
By examining both parameterized complexity and approximation algorithms, we pinpointed treewidth, thickness and density as key parameters that influence computational complexity. These theoretical results suggest that solving the UD-MIS problem is easier for instances with low density or small treewidth. We also highlight that complexity theory does not shed light on the differences between the MIS and weighted MIS in terms of running times on UD graphs, i.e., weights do not appear to influence the theoretical complexity. Specifically, the previously discussed results also extend to the MWIS problem. For PTAS, see Ref.~\cite{th:weightedPTASbonamy2021eptas}. Regarding treewidth, the solving algorithm follows trivially from Ref.~\cite{lipton1977applications}. For EPTAS in bounded-density graphs, refer to Ref.~\cite{th:geometricThesis}. Finally, for sub-exponential algorithms, we point to an upcoming work~\cite{Pierretoappear}.
Finally, the parameterized complexity of UD-MWIS with respect to thickness remains FPT, since bounded thickness implies bounded treewidth, allowing the previously mentioned treewidth-based algorithm to be applied directly.
On the other side, weights can impact experimental outcomes, despite no significant difference being expected in the asymptotic regime. For the reasons above, in Section~\ref{sec:numerics} we will also explore how changing weights in the MWIS can result in harder instances and how much this could affect the solving time.

\section{Benchmarking parametrized complexity with classical solvers}
\label{sec:numerics}

In the previous section, we extracted from theoretical results some graph parameters that are not related just to size, but to more subtle structures of the graphs as well, including density, thickness and treewidth. In this section, we want to verify numerically if these parameters can act as hardness parameters, and quantify the hardness increase with respect to each one.

\subsection{Justification of the classical solver and the dataset}
\label{sec:branchandbound}

Assessing the hardness of a problem requires testing a variety of classical optimizers. In Ref.~\cite{andrist2023hardness}, the authors evaluated several classical solvers for MIS on KL graphs, ranging from heuristic approaches to exact methods. Similar to their approach, our study aims to understand how changes in a hardness parameter impact the solving time of a classical solver. For this purpose, we chose to use CPLEX~\cite{cplex2024}, a commercial off-the-shelf solver known for guaranteeing optimality in solving the MIS problem. It utilizes a branch-and-cut algorithm, which breaks large optimization problems into smaller, more manageable sub-problems.

We evaluate the solver performance by tracking the number of steps it takes to reach a solution, measured in \textit{deterministic ticks}. A tick represents a memory access made by the solver during the optimization process, effectively counting the computational steps taken towards a solution. This metric ensures reproducibility across different computing environments, unlike wall clock time, which can fluctuate due to system load or hardware differences. By using ticks, we obtain a stable and consistent measure of computational effort.

In our setup, the solver operated at approximately $300$ ticks per second on a 32-core CPU (Dual AMD Rome 7742, 128 cores total, 2.25 GHz). Additionally, we retained CPLEX pre-solving methods, which use advanced heuristics to prune the search space. This choice preserves its `off-the-shelf' quality, ensuring that our results reflect its standard performance. It is important to emphasize that we did not fine-tune the solver for this specific problem. As a result, our reported runtimes serve as upper bounds on the best possible solving times achievable by classical solvers. However, our primary focus concerns how runtime scales across different parameter ranges, a trend effectively captured by the ticks metric.

We generate a dataset of graphs with varying density by selecting a pre-defined, regular triangular layout as described in App.~\ref{sec:generate_instances}. Our choice is motivated by two key factors. The geometry of the triangular layout minimizes the impact of long-range interactions in the Ising Hamiltonian, reducing next-nearest-neighbor contributions to $3.7\%$ of the nearest-neighbor interaction, compared to $6.4\%$ in King’s layout. This ensures a more accurate embedding of UD graphs, as unconnected pairs of nodes experience minimal interaction. Additionally, regular triangular layouts are known to exhibit classical magnetic frustration~\cite{wannier1950antiferromagnetism}, a regime that appears particularly challenging for classical devices to handle.

\subsection{Impact of density}

In Fig.~\ref{fig:fig1}c, we show how both the density and the instance size 
affect the CPLEX computation time. The horizontal axis corresponds to the 
density \(\rho\), while the vertical axis indicates the number of vertices \(N\). 
The color scale reflects \(\mathrm{TTS}_c\) (the time to solution), shown on a 
logarithmic scale from \(10^{-1}\) to \(10^3\). Lighter shades indicate shorter 
runtimes, whereas darker shades signify longer runtimes.

These native instances are generated via the density model described in 
Appendix~\ref{sec:generate_instances}. At lower densities or smaller sizes, the 
problem is relatively straightforward to solve, leading to shorter runtimes. In 
contrast, small but dense instances can become notably more challenging, 
sometimes surpassing the difficulty of larger, yet sparser instances. This 
reflects how the sheer number of possible edges in a dense graph can heavily 
constrain the solver. Nevertheless, at \(\rho = 1\) (corresponding to a fully 
connected sub-layout), the problem appears to simplify again, likely due to the 
high degree of regularity in the resulting graph.

In Appendix~\ref{app:density}, we further explore how artificially increasing 
the density of the embedded graphs affects their runtime. Although such graphs 
might not be directly realizable on hardware in their current form, one could 
achieve higher densities by increasing the Rydberg blockade radius or reducing 
the minimum spacing between atoms. Indeed, a recent experiment has demonstrated 
how atoms can be brought arbitrarily close by using optical tweezers 
\cite{hwang2024impact}, providing a potential pathway for probing these dense 
regimes in practice.

\subsection{Impact of treewidth}

\begin{figure}
    \centering
\includegraphics[width=\linewidth]{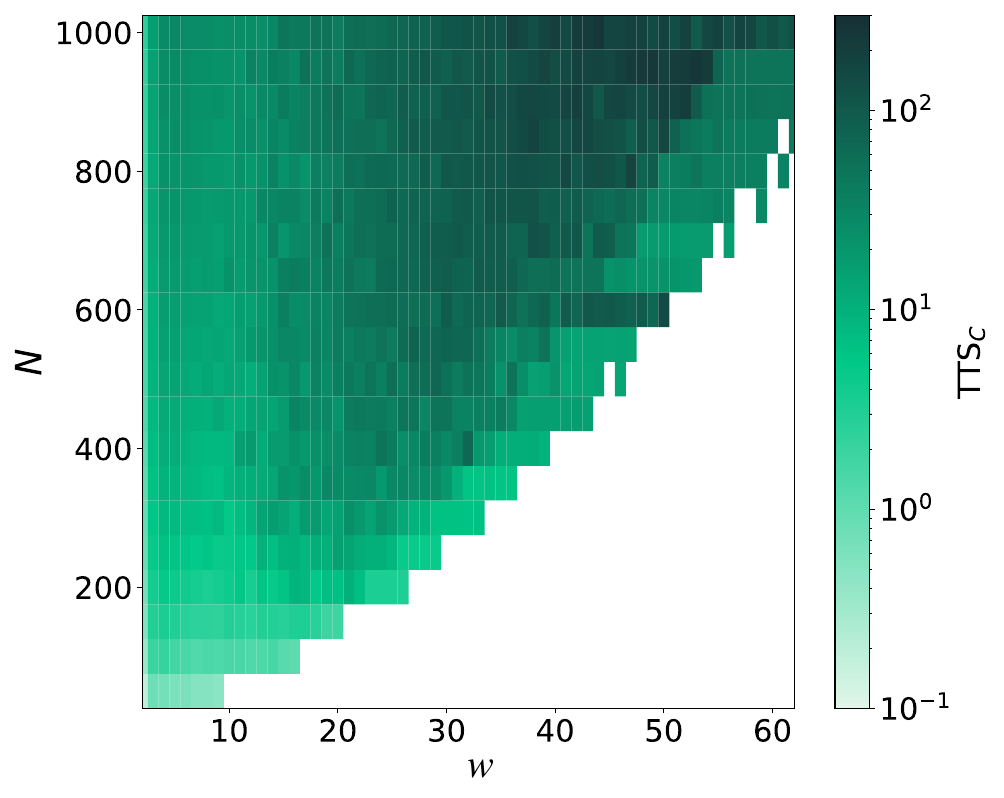}
    \caption{Treewidth $w$ as a hardness parameter for MIS in UD graphs generated according to App.~\ref{sec:generate_instances}, as in Fig.~\ref{fig:fig1}$(c)$. The treewidth of the graphs is approximated by means of \textit{networkx} library~\cite{hagberg2008exploring}.}
    \label{fig:tw_TTS}
\end{figure}

In Fig.~\ref{fig:tw_TTS}, we illustrate how the approximate treewidth \(w\) 
affects the CPLEX runtime \(\mathrm{TTS}_c\), plotted on a logarithmic color 
scale, as a function of graph size \(N\). Here, \(w\) is estimated using the 
\textit{Min-Fill} heuristic in the \texttt{networkx} library 
\cite{hagberg2008exploring}; while exact treewidth algorithms were tested, they 
proved intractable for large or dense graphs.

Overall, the figure exhibits parallels to the density‐based analysis: low 
treewidth or small graphs yield shorter runtimes (lighter shades), whereas 
instances of moderate to large \(w\) become increasingly challenging (darker 
shades). Interestingly, at very high treewidth values, which typically arise 
in dense structures, the solver’s runtime decreases again. This 
behavior mirrors the “regular‐layout” effect seen in fully dense graphs: 
uniformity in the connectivity can reduce the complexity of the constraint 
search space. 

Moreover, treewidth tends to be positively correlated with graph density, 
reaching its maximum for clique‐like graphs. In Appendix~\ref{app:UD_to_ER}, 
we investigate how randomly altering edges—thereby changing the treewidth 
while keeping density fixed—affects the runtime. Those experiments confirm 
that larger treewidth alone (independently of density) can drive up the 
complexity of solving these native quantum‐embeddable instances, highlighting 
treewidth as a key factor in runtime variability.

\subsection{Impact of weight}
\label{subsec:Impact of weight}
\begin{figure*}
    \centering 
\includegraphics[width=0.9\linewidth]{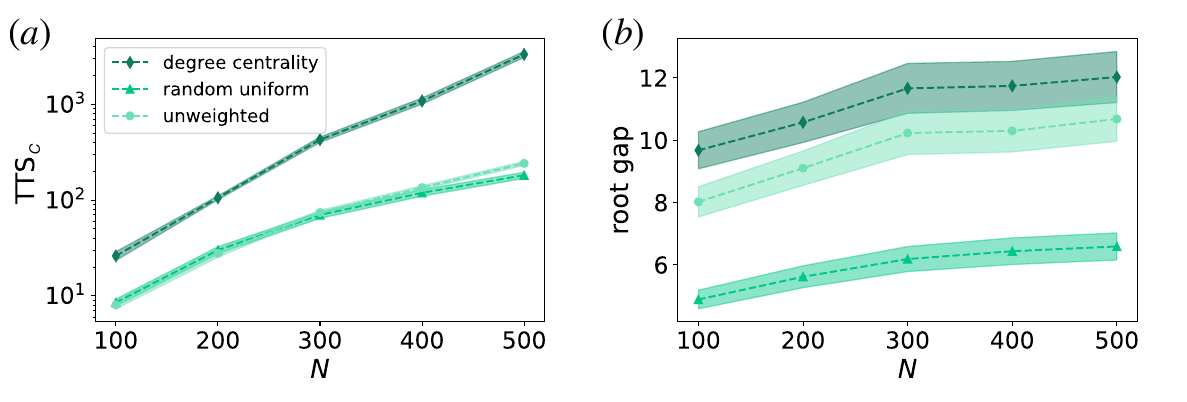}
    \caption{$(a)$ Resolution time according to the CPLEX solver and $(b)$ root gap in the case of native instances according to the sampling described in Fig.~\ref{fig:fig1} and App.~\ref{sec:generate_instances}. The results averaged over 50 repetitions, and the shadowed areas represent the errors on the mean.}
    \label{fig:weight_study}
\end{figure*}
\label{sec:impactofweights}
In this section, we present a constructive method for assigning weights to the nodes of UD graphs, designed to make the resolution of the Maximum Weighted Independent Set (MWIS) problem more challenging for classical solvers.

While this section presents the most efficient among the ones explored, details of the remaining methods are provided in Appendix~\ref{appendix:weight_methods}.

To design challenging MWIS instances, we assign weights to nodes based on their degree centrality. The degree centrality captures the fraction of nodes in the graph to which a given node is directly connected, and it is defined as:
\begin{equation}
    C_D(v) = \frac{\text{deg}(v)}{n - 1}
\end{equation}
where \(\text{deg}(v)\) is the degree of the node \(v\), and \(n\) is the total number of nodes in the graph. The weights are then assigned proportionally to the normalized degree centrality as follows:
\begin{equation}
    w(v) = 1 + \bar{\delta} \left( \frac{C_D(v)}{\max(C_D(u))} \right) \quad \forall u \in V.
\end{equation}
Here, $\bar{\delta}$ is a constant that we empirically choose as $\bar{\delta}=1000$.
To thoroughly evaluate the computational complexity introduced by this method, we benchmark it against the following baseline approaches: \textit{(i)} unweighted graphs, where all node weights are set to 1, and \textit{(ii)} random uniform weights, where each node is assigned a weight drawn uniformly at random from the interval \([0.1, \bar{\delta}]\). The experiments were conducted on random sparse UD graphs natively implementable in neutral atom quantum processors. 
We generated graphs with varying sizes (\(100, 200, 300, 400, 500\) nodes). For each size, the results were averaged over $50$ random unique instances. 
The computational time was measured in terms of deterministic ticks~\cite{cplex2024} (as in  Fig.~\ref{fig:fig1}). The results are illustrated in Fig.~\ref{fig:weight_study}.

As shown in Fig.\ref{fig:weight_study}$(a)$ the computation time varies significantly across the different weighting schemes, reflecting their impact on the complexity of solving MWIS instances. The degree centrality-based weighting scheme shows a much faster growth in computation time as the graph size increases. For graphs with 500 nodes, the computation time for degree centrality weights is, on average, approximately 10 times longer than for random uniform weights, and up to 16 times longer in some cases. In contrast, the unweighted and random uniform weighting schemes demonstrate a more gradual increase in computation time as the graph size increases. Overall, the trends observed in this figure underline the strong influence of the weighting scheme on the computational performance of MWIS solvers, with degree centrality weights consistently leading to higher computational complexity.

To explain this behavior, we employ the root gap and show it in Fig.~\ref{fig:weight_study}$(b)$.
The root gap represents the difference in the objective value between the solution to the linear relaxation at the root node and the optimal integer solution~\cite{lawler1966branch}. It is an indicator of difficulty because a larger gap usually means the branch-and-cut algorithm needs to perform more branching and add additional cutting planes to close the gap and reach the optimal solution.
For degree centrality weights, the root gap consistently remains the highest across all graph sizes, exceeding $70\%$ for larger graphs. This large root gap signifies a greater challenge in closing the gap between the linear relaxation and the optimal integer solution, thereby necessitating extensive branching and additional cutting planes, which significantly increases computation time. In contrast, the random uniform weights exhibit the lowest root gap, remaining consistently below $30\%$ for all graph sizes, which correlates with the shorter computation times observed in Fig.~\ref{fig:weight_study}$(a)$. The unweighted scheme falls in between, with root gaps typically ranging from $50\%$ to $60\%$, suggesting moderate difficulty. The trends observed reinforce that the weighting scheme directly impacts the computational effort required, with degree centrality-based weights being notably more challenging to solve.
These numerical observations help to explain the increased challenges associated with this weighting scheme. A formal theoretical analysis is beyond the scope of this paper, but is part of ongoing work~\cite{Pierretoappear}.

\section{Benchmark on native instances with neutral atoms }
\label{sec:experiments}
The goal of this section is exploring the performances of neutral atoms quantum platforms for the MIS problem, keeping in mind
that we are interested in investigating native instances, i.e., graphs that are compatible with the inherent physical properties of the system. We start by introducing the quantum algorithm used for tackling the problem. Then, we study the quality of the solution, in both numerics and experiments, as a function of the size (number of nodes) and density of the graph. Finally, we investigate the impact of weights on the nodes according to the discussion in Sec.~\ref{sec:impactofweights}.

\begin{figure*}
\centering
\includegraphics[width=\linewidth]{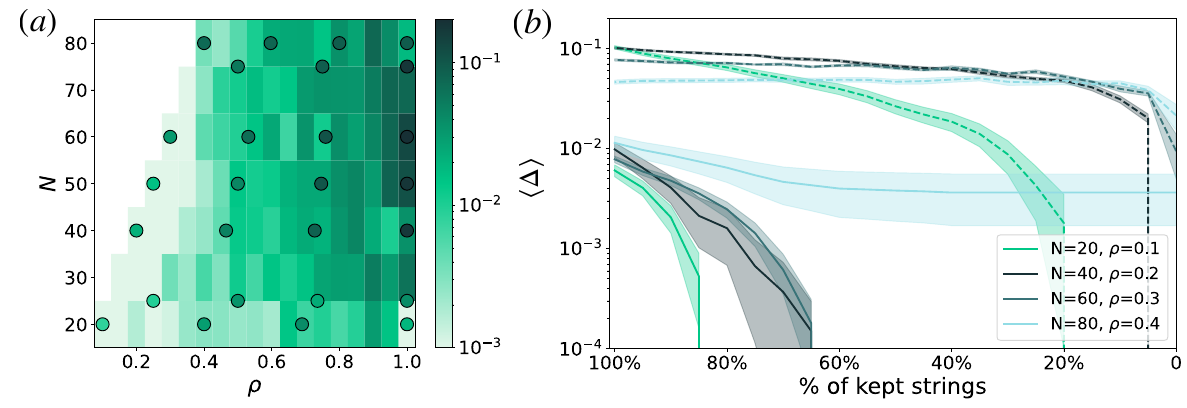}
    \caption{$(a)$ Contour plot of the average gap to solution as function of size and density. $(b)$ Average gap to solution as a function of the percentage of bitstrings that are employed for calculating the gap, for different system sizes $N$ and densities $\rho$. The results averaged over 10 repetitions, and the shadowed areas represent the errors on the mean.}
    \label{fig:quantum_MIS}
\end{figure*}

\subsection{MWIS solving with neutral atoms}
We employ a neutral-atom based QPU made of single $^{87}$Rb atoms trapped in arrays of optical tweezers~\cite{barredo2018synthetic,Nogrette14,browaeys2020many,henriet2020quantum,Morgado2021}. Qubits are encoded onto the ground state $\ket{0}=\ket{5S_{1/2},F=2,m_F=2}$ and the Rydberg state $\ket{1}=\ket{60S_{1/2}, m_J=1/2}$ of the atoms.
The dynamics of a collection of $N$ qubits at positions $\boldsymbol{r}$ are governed by the following Hamiltonian $\hat H(t)=\hat H_{\rm drive}(t)+\hat H_{\rm cost}(t)$, where:
\begin{equation}
\begin{aligned}
&\hat H_{\rm drive}(t)/\hbar=\frac{\Omega(t)}{2}\sum_{i=1}^N\hat\sigma_i^x,\\
&\hat H_{\rm cost}(t)/\hbar=-\delta(t)\sum_{i=1}^N\epsilon_i\hat n_i+\sum_{i<j}U(|r_i-r_j|)\hat n_i \hat n_j.\\
\end{aligned}
\end{equation}
Here, $\hat{\sigma}_i^\alpha$
are Pauli matrices and $\hat n_i=\ket{1}_i\bra{1}_i$. The two time-dependent control fields are the Rabi frequency $\Omega(t)$ and the detuning $\delta(t)$. $U(d)=C_6/|r_i-r_j|^{6}$ is the position-dependent interaction function, with $C_6/h=138~$GHz$\cdot \mu {\rm m}^6$ for the Rydberg state considered \cite{Beguin_2013,ibali2017}.  The driving part of the Hamiltonian provides a way to excite the atoms, and the cost part a way to encode the optimization problem.
In particular, for a graph $\Graph$, the MWIS cost function can be encoded as an operator of the form:
\begin{equation}
    C_\Graph(\hat n)=\hat C_\Graph=-\sum_{i\in\vertices}w_i\hat n_i+\alpha \sum_{i<j}\hat n_i \hat n_j.
\end{equation}
This is equivalent to Eq.~\eqref{eq:mis_class}, where the binary variable $x_i$ is substituted by the number operator $n_i$.
The nodes are weighted with $w_i\in[0,1]$, and edges between adjacent nodes carry a uniform weight of $\alpha>0$. In the case of the weights $w_i$ being uniformly $1$, the problem to solve becomes an MIS. The ground state of $\hat C_\Graph$ manifests as a coherent superposition of MWIS classical product states. Such an operator finds correspondence with the cost Hamiltonian $\hat H_{\rm cost}$ (with $\delta>0$) of $N$ atoms well-positioned such that:
\begin{equation}
        \hat H_{\rm cost}/\hbar \approx \delta\left(-\sum_{i=1}^N w_i\hat n_i+\frac{U(r_b)}{\delta} \sum_{i<j}\hat n_i \hat n_j\right) =\delta \hat C_\Graph
\end{equation}
where we take $\epsilon_i=w_i$ and the ratio between the nearest-neighbors interaction $U(r_b)$ and the detuning encodes the uniform edge weight $\alpha$. 
By preparing the ground state of $\hat H_{\rm cost}$, and subsequently sampling it $n_{\rm shots}$ times, it becomes possible to measure not just one, but all the MWISs of $\Graph$ as they are degenerate.
The procedure employed for solving the MWIS problem is the following.
A random native instance (graph) at fixed value of density and number of nodes is generated. The quantum state is initialized as $\ket{\psi}=\ket{0}^{\otimes N}$ and driven towards the correct final target state through a slow evolution according to the paradigm of adiabatic quantum computation~\cite{farhi2014quantum}.
The generation of the graphs and the particular annealing schedule used are shown in App.~\ref{app:quantum}.

\begin{figure*}
\centering
\includegraphics[width=\linewidth]{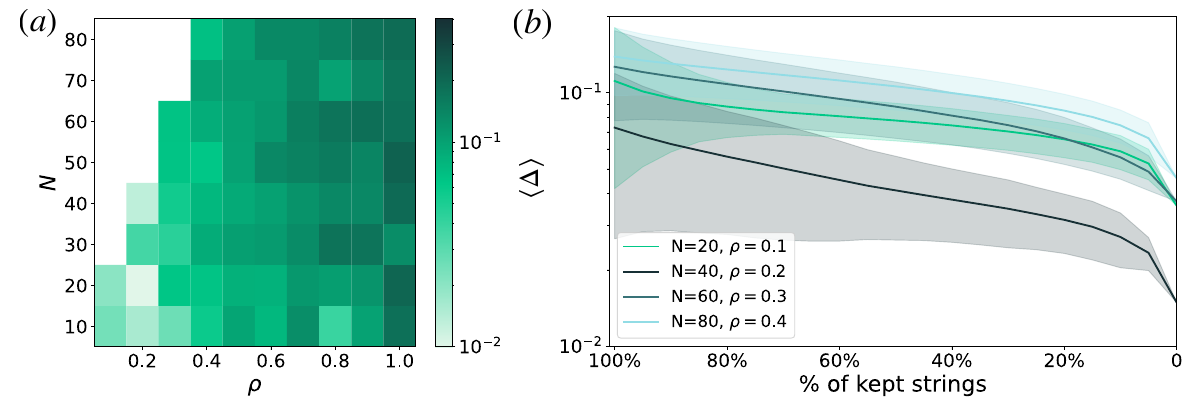}
\caption {$(a)$ color plot of the average gap to optimal solution as a function of size $N$ and density $\rho$. $(b)$ Average gap to solution as a function of the percentage of bitstrings that are employed for calculating the gap, for different system sizes $N$ and densities $\rho$.} 

    \label{fig:quantum_MWIS}
\end{figure*}

\subsection{MIS problem: numerical and experimental results}

In Fig.~\ref{fig:quantum_MIS} we plot the numerical and experimental results of the solution of the MIS problem employing a neutral atom quantum platform. In Fig.~\ref{fig:quantum_MIS}$(a)$ we show a color plot of the average (over 10 instances) gap to solution
$\langle \Delta \rangle=\langle \left| \frac{C_\Graph -S_\Graph}{S_\Graph} \right|\rangle$ where $S_\Graph$ is the size of the MIS problem solution on the sampled graph $\Graph$ .
The simulations are performed by means of tensor networks~\cite{emu-mps}.
We observe that the gap increases as the system size and the density grow, testifying the increase in hardness of the problem and the impossibility of finding reliably the optimal solution with the employed annealing schedule. The colored dots represent the experimental results that are obtained after error mitigation and processing. 
We observe that the experimental results show a similar trend compared to the simulations, but the performances appear slightly worse, due to experimental imperfections and noise.
The general quality of the distributions, 
simulated or experimentally obtained, might be better summarized
by looking at the truncated
average gap as in Fig.~\ref{fig:quantum_MIS}$(b)$.
This figure shows how the quality of the solution improves when the worst cost sampled bitstrings are discarded more and more.
The simulations (solid lines) approach rapidly zero, despite the case $N=80$ where the quality of the solution plateaus even when keeping only the best cost strings, meaning that the quantum annealing was not effective to find the MIS in all the instances explored.
This is particularly interesting in light of the experiments that are shown in the same colors using dashed lines. We observe that the average gap dramatically improves when retaining only a few percent of the sampled bitstrings, shedding light on the fact that we are able to sample the strings close to the MIS.
In particular, in the case $N=80$, the best cost bitstrings have an average gap that is compatible with the simulations results within error bars.

\subsection{MWIS: impact of weights}

The results of the simulations for the weighted MIS problem are presented in Fig.~\ref{fig:quantum_MWIS}. Ranging from $N=10$ to $N=80$ nodes and densities from $\rho=0.1$ to $\rho=1.0$, a color plot of the gap to optimality is exhibited in Fig.~\ref{fig:quantum_MWIS}$(a)$. Whilst lighter shades signify smaller gaps, darker ones are attributed to the higher values. The general tendency to have larger gaps is manifest while both $N$ and $\rho$ increase. Although some of the maximal gap values correspond to the lower or medium number of nodes, they are found at the higher density values. It is worth pointing out that the largest gaps conform with the largest density, $\rho=1.0$, implying its role in making a graph more difficult to solve. When consulting  the points regarding the top 25\% gaps, one can see that all of them correspond to densities ranging from 0.7 to 1.0, and about 70\% of them to 50 nodes or more. 

In Fig.~\ref{fig:quantum_MWIS}$(b)$, for some specific ($N$, $\rho$), the average gap is shown as a function of the percentage of the bitstrings with best costs  that are kept to calculate it. As expected, the values are lower for smaller ($N$, $\rho$), and also as the curves advance towards the lower percentages. For 100\% of the bitstrings, the gap varies from 0.073 ($N=40$ and $\rho=0.2$) to 0.138 ($N=80$ and $\rho=0.4$), while for the 20\% best bitstrings, they ranged from 0.032 ($N=40$ and $\rho=0.2$) to 0.085 ($N=80$ and $\rho=0.4$). 

The average gap to solution values in Fig.~\ref{fig:quantum_MWIS}$(a)$ are higher than its non-weighted counterpart, more flagrantly for the lowest densities, even for the bigger number of nodes. It is also worth noticing the difference in the scales of the two plots when comparing them: while here the gap values go as low as $10^{-2}$, in the previous case (MIS) they fall to $10^{-3}$, evidencing how weighting the graphs represent an increase in the difficulty to find an optimal solution. 
For the Fig.~\ref{fig:quantum_MWIS}$(b)$, in comparison with Fig.~\ref{fig:quantum_MIS}$(b)$, it is patent the increase in the gap values, even when a bigger percentage of the worst bitstrings is discarded: in none of the cases a rapid drop in the value is seen, except when the percentage is very close to 0\%. That testifies for the hardness added along with the weights: the gaps for the MIS Fig.~\ref{fig:quantum_MIS}$(b)$ start one order of magnitude below the MWIS and reach two orders of magnitude lower whereas, for the latter, they don't go under $10^{-2}$. That means one has to take into account just a little top fraction of the best bitstrings, and it still will not be a result as good as the non-weighted case.

\section{Conclusion and perspectives}

In this paper, we have provided a theoretical foundation for understanding the complexity of the UD-MIS problem and evaluated the performance of neutral-atom quantum hardware through a rigorous benchmark. Our results indicate that both quantum and classical methods exhibit qualitatively similar behavior with respect to hardness parameters. However, for small problem instances, classical solvers currently outperform quantum ones by approximately three orders of magnitude, primarily due to the low repetition rate of quantum hardware. While quantum state preparation—the most computationally challenging step—occurs at the MHz scale, the extraction of results via projective measurements remains comparatively slow. The repetition rate is currently limited to a few Hz due to constraints in atom loading, spatial arrangement, and imaging~\cite{young2020half,gyger2024continuous}. This remains six orders of magnitude below the fundamental limit of neutral atoms, highlighting a major engineering challenge. Addressing these bottlenecks requires improvements in hardware design. Techniques such as atom reservoir loading, parallel transport, and optimized assembly algorithms could significantly boost the repetition rate. Additionally, fast imaging using high numerical aperture optics, cryogenic avalanche detectors, or cavity-assisted detection methods could push this rate towards 100 Hz or even the kHz range~\cite{Cimring_2023,Bergschneider_2018, Buzulutskov_2012}.

Another key step toward achieving parity with classical solvers is scaling up instance sizes to the thousands, where classical runtimes become noticeably longer. Benchmarking larger problems is crucial for understanding the scaling behavior of quantum approaches, particularly for neutral atom architectures. Recent experimental progress—demonstrating neutral atom arrays with up to 6,100 qubits~\cite{manetsch2024tweezer}—suggests a promising pathway for scalability.

Beyond hardware improvements, finding harder problem instances is equally important. While we have focused on exact solutions, recent work suggests that quantum annealing may exhibit a scaling advantage in approximate optimization~\cite{bauza2024scaling}. However, this advantage only becomes meaningful if the underlying problem instances are sufficiently hard, which we have characterized for neutral atoms in this work. While the UD-MIS problem is NP-hard in the worst case, its average-case hardness is relatively low, and strong approximation schemes exist. Despite our efforts to extract maximum hardness from unit-disk graphs, we encourage the community to seek embeddings of harder native problem instances. This could significantly raise the bar for classical solvers and broaden the scope of combinatorial optimization problems accessible to quantum methods. 
A promising direction in this regard is Hamiltonian engineering, where neutral atoms can be programmed to realize complex, non-local interactions. Recent work has demonstrated the feasibility of tailored interactions~\cite{scholl2022microwave}, 3D arrays~\cite{barredo2018synthetic,lee2016three} and dynamical connectivity~\cite{bluvstein2022quantum}, opening up new possibilities for quantum optimization that classical solvers struggle to address.

Overall, while a clear quantum advantage in combinatorial optimization with neutral atoms has yet to be demonstrated, the pathway forward is becoming increasingly well-defined. Key advancements lie in both hardware engineering and problem encoding.  With continued progress, we are optimistic that neutral atom quantum processors will not only reach computational regimes that rival classical solvers but will also unlock entirely new capabilities for tackling complex optimization problems.

\section*{Acknowledgments}

We thank Alexandre Dauphin, Helmut Katzgraber, Salavatore Mandrà and Louis-Paul Henry for enlightening discussions. We are also grateful to the Hardware and Cloud team from Pasqal. VV is grateful to Sergi Julià-Farré for valuable insights on the tensor networks simulations. CD would also like to thank Marie Wakim for making the plots more fashionable.

\bibliographystyle{unsrtnat}
\bibliography{refs_optim_with_dois}

\begin{thebibliography}{108}
\providecommand{\natexlab}[1]{#1}
\providecommand{\url}[1]{\texttt{#1}}
\expandafter\ifx\csname urlstyle\endcsname\relax
  \providecommand{\doi}[1]{doi: #1}\else
  \providecommand{\doi}{doi: \begingroup \urlstyle{rm}\Url}\fi

\bibitem[Paschos(2014)]{paschos2014applications}
Vangelis~Th Paschos.
\newblock \emph{Applications of combinatorial optimization}.
\newblock John Wiley \& Sons, 2014.
\newblock \doi{10.1002/9781118600283}.

\bibitem[Resende(2003)]{resende2003combinatorial}
Mauricio~GC Resende.
\newblock Combinatorial optimization in telecommunications.
\newblock \emph{Optimization and industry: new frontiers}, pages 59--112, 2003.
\newblock \doi{10.1007/978-1-4613-0233-9_4}.

\bibitem[Du et~al.(2019)Du, Zheng, and Ouyang]{du2019real}
Gang Du, Luyao Zheng, and Xiaoling Ouyang.
\newblock Real-time scheduling optimization considering the unexpected events in home health care.
\newblock \emph{Journal of Combinatorial Optimization}, 37:\penalty0 196--220, 2019.
\newblock \doi{10.1007/s10878-017-0220-3}.

\bibitem[Ali et~al.(2015)Ali, Adewumi, Blamah, and Falowo]{ali2015mathematical}
M~Montaz Ali, Aderemi~O Adewumi, Nachamada Blamah, and Olabisi Falowo.
\newblock Mathematical modeling and optimization of industrial problems.
\newblock \emph{Journal of Applied Mathematics}, 2015, 2015.
\newblock \doi{10.1155/2015/438471}.

\bibitem[Shor(1995)]{Shor1995PolynomialTimeAF}
Peter~W. Shor.
\newblock Polynomial-time algorithms for prime factorization and discrete logarithms on a quantum computer.
\newblock \emph{SIAM Rev.}, 41:\penalty0 303--332, 1995.
\newblock URL \url{https://api.semanticscholar.org/CorpusID:2337707}.

\bibitem[Farhi et~al.(2014{\natexlab{a}})Farhi, Goldstone, and Gutmann]{farhi2014quantum}
Edward Farhi, Jeffrey Goldstone, and Sam Gutmann.
\newblock A quantum approximate optimization algorithm.
\newblock \emph{arXiv preprint arXiv:1411.4028}, 2014{\natexlab{a}}.
\newblock \doi{10.1103/physreva.111.012427}.

\bibitem[Abbas et~al.(2023)Abbas, Ambainis, Augustino, B{\"a}rtschi, Buhrman, Coffrin, Cortiana, Dunjko, Egger, Elmegreen, et~al.]{abbas2023quantum}
Amira Abbas, Andris Ambainis, Brandon Augustino, Andreas B{\"a}rtschi, Harry Buhrman, Carleton Coffrin, Giorgio Cortiana, Vedran Dunjko, Daniel~J Egger, Bruce~G Elmegreen, et~al.
\newblock Quantum optimization: Potential, challenges, and the path forward.
\newblock \emph{arXiv preprint arXiv:2312.02279}, 2023.

\bibitem[Tasseff et~al.(2024)Tasseff, Albash, Morrell, Vuffray, Lokhov, Misra, and Coffrin]{tasseff2024emerging}
Byron Tasseff, Tameem Albash, Zachary Morrell, Marc Vuffray, Andrey~Y Lokhov, Sidhant Misra, and Carleton Coffrin.
\newblock On the emerging potential of quantum annealing hardware for combinatorial optimization.
\newblock \emph{Journal of Heuristics}, 30\penalty0 (5):\penalty0 325--358, 2024.
\newblock \doi{10.1007/s10732-024-09530-5}.

\bibitem[Ebadi et~al.(2021)Ebadi, Wang, Levine, Keesling, Semeghini, Omran, Bluvstein, Samajdar, Pichler, Ho, et~al.]{ebadi2021quantum}
Sepehr Ebadi, Tout~T Wang, Harry Levine, Alexander Keesling, Giulia Semeghini, Ahmed Omran, Dolev Bluvstein, Rhine Samajdar, Hannes Pichler, Wen~Wei Ho, et~al.
\newblock Quantum phases of matter on a 256-atom programmable quantum simulator.
\newblock \emph{Nature}, 595\penalty0 (7866):\penalty0 227--232, 2021.
\newblock \doi{10.1038/s41586-021-03582-4}.

\bibitem[Dalyac et~al.(2024)Dalyac, Leclerc, Vignoli, Djellabi, Coelho, Ximenez, Dareau, Dreon, Elfving, Signoles, et~al.]{dalyac2024graph}
Constantin Dalyac, Lucas Leclerc, Louis Vignoli, Mehdi Djellabi, Wesley da~Silva Coelho, Bruno Ximenez, Alexandre Dareau, Davide Dreon, VIncent~E Elfving, Adrien Signoles, et~al.
\newblock Graph algorithms with neutral atom quantum processors.
\newblock \emph{arXiv preprint arXiv:2403.11931}, 2024.
\newblock \doi{10.1140/epja/s10050-024-01385-5}.

\bibitem[Dalyac et~al.(2021)Dalyac, Henriet, Jeandel, Lechner, Perdrix, Porcheron, and Veshchezerova]{dalyac2021qualifying}
Constantin Dalyac, Lo{\"\i}c Henriet, Emmanuel Jeandel, Wolfgang Lechner, Simon Perdrix, Marc Porcheron, and Margarita Veshchezerova.
\newblock Qualifying quantum approaches for hard industrial optimization problems. a case study in the field of smart-charging of electric vehicles.
\newblock \emph{EPJ Quantum Technology}, 8\penalty0 (1):\penalty0 12, 2021.
\newblock \doi{10.1140/epjqt/s40507-021-00100-3}.

\bibitem[Byun et~al.(2022{\natexlab{a}})Byun, Kim, and Ahn]{byun2022finding}
Andrew Byun, Minhyuk Kim, and Jaewook Ahn.
\newblock Finding the maximum independent sets of platonic graphs using rydberg atoms.
\newblock \emph{PRX Quantum}, 3\penalty0 (3):\penalty0 030305, 2022{\natexlab{a}}.
\newblock \doi{10.1103/prxquantum.3.030305}.

\bibitem[Leclerc et~al.(2024)Leclerc, Dalyac, Bendotti, Griset, Mikael, and Henriet]{leclerc2024implementingtransferableannealingprotocols}
Lucas Leclerc, Constantin Dalyac, Pascale Bendotti, Rodolphe Griset, Joseph Mikael, and Loïc Henriet.
\newblock Implementing transferable annealing protocols for combinatorial optimisation on neutral atom quantum processors: a case study on smart-charging of electric vehicles, 2024.
\newblock URL \url{https://arxiv.org/abs/2411.16656}.

\bibitem[Lucas(2014)]{lucas2014ising}
Andrew Lucas.
\newblock Ising formulations of many np problems.
\newblock \emph{Frontiers in physics}, 2:\penalty0 5, 2014.
\newblock \doi{10.3389/fphy.2014.00005}.

\bibitem[Pichler et~al.(2018)Pichler, Wang, Zhou, Choi, and Lukin]{pichler2018quantum}
Hannes Pichler, Sheng-Tao Wang, Leo Zhou, Soonwon Choi, and Mikhail~D Lukin.
\newblock Quantum optimization for maximum independent set using rydberg atom arrays.
\newblock \emph{arXiv preprint arXiv:1808.10816}, 2018.
\newblock \doi{10.26226/m.6275705c66d5dcf63a3115ad}.

\bibitem[Lukin et~al.(2001)Lukin, Fleischhauer, Cote, Duan, Jaksch, Cirac, and Zoller]{lukin2001dipole}
Mikhail~D Lukin, Michael Fleischhauer, Robin Cote, LuMing Duan, Dieter Jaksch, J~Ignacio Cirac, and Peter Zoller.
\newblock Dipole blockade and quantum information processing in mesoscopic atomic ensembles.
\newblock \emph{Physical review letters}, 87\penalty0 (3):\penalty0 037901, 2001.
\newblock \doi{10.1103/physrevlett.87.037901}.

\bibitem[Browaeys and Lahaye(2020)]{browaeys2020many}
Antoine Browaeys and Thierry Lahaye.
\newblock Many-body physics with individually controlled rydberg atoms.
\newblock \emph{Nature Physics}, 16\penalty0 (2):\penalty0 132--142, 2020.
\newblock \doi{10.1038/s41567-019-0733-z}.

\bibitem[Saffman et~al.(2010)Saffman, Walker, and M{\o}lmer]{saffman2010quantum}
Mark Saffman, Thad~G Walker, and Klaus M{\o}lmer.
\newblock Quantum information with rydberg atoms.
\newblock \emph{Reviews of modern physics}, 82\penalty0 (3):\penalty0 2313--2363, 2010.
\newblock \doi{10.1117/12.2519368}.

\bibitem[Urban et~al.(2009)Urban, Johnson, Henage, Isenhower, Yavuz, Walker, and Saffman]{urban2009observation}
E~Urban, Todd~A Johnson, T~Henage, L~Isenhower, DD~Yavuz, TG~Walker, and M~Saffman.
\newblock Observation of rydberg blockade between two atoms.
\newblock \emph{Nature Physics}, 5\penalty0 (2):\penalty0 110--114, 2009.
\newblock \doi{10.1038/nphys1178}.

\bibitem[Labuhn et~al.(2016)Labuhn, Barredo, Ravets, De~L{\'e}s{\'e}leuc, Macr{\`\i}, Lahaye, and Browaeys]{labuhn2016tunable}
Henning Labuhn, Daniel Barredo, Sylvain Ravets, Sylvain De~L{\'e}s{\'e}leuc, Tommaso Macr{\`\i}, Thierry Lahaye, and Antoine Browaeys.
\newblock Tunable two-dimensional arrays of single rydberg atoms for realizing quantum ising models.
\newblock \emph{Nature}, 534\penalty0 (7609):\penalty0 667--670, 2016.
\newblock \doi{10.1038/nature18274}.

\bibitem[Note1()]{Note1}
Note1.
\newblock A unit-disk graph is defined by the positions of nodes, where an edge exists between two nodes if the distance between them is less than a constant \( r \), typically set to \( r = 1 \).

\bibitem[Kim et~al.(2022)Kim, Kim, Hwang, Moon, and Ahn]{kim2022rydberg}
Minhyuk Kim, Kangheun Kim, Jaeyong Hwang, Eun-Gook Moon, and Jaewook Ahn.
\newblock Rydberg quantum wires for maximum independent set problems.
\newblock \emph{Nature Physics}, 18\penalty0 (7):\penalty0 755--759, 2022.
\newblock \doi{10.1038/s41567-022-01629-5}.

\bibitem[Byun et~al.(2022{\natexlab{b}})Byun, Kim, and Ahn]{byun2022quantum}
Andrew Byun, Minhyuk Kim, and Jaewook Ahn.
\newblock Quantum simulation of ising spins on platonic graphs.
\newblock \emph{arXiv preprint arXiv:2203.01541}, 2022{\natexlab{b}}.
\newblock \doi{10.1016/0550-3213(94)90033-7}.

\bibitem[Dalyac et~al.(2023)Dalyac, Henry, Kim, Ahn, and Henriet]{dalyac2023exploring}
Constantin Dalyac, Louis-Paul Henry, Minhyuk Kim, Jaewook Ahn, and Lo{\"\i}c Henriet.
\newblock Exploring the impact of graph locality for the resolution of the maximum-independent-set problem with neutral atom devices.
\newblock \emph{Physical Review A}, 108\penalty0 (5):\penalty0 052423, 2023.
\newblock \doi{10.1103/physreva.108.052423}.

\bibitem[Nguyen et~al.(2023)Nguyen, Liu, Wurtz, Lukin, Wang, and Pichler]{nguyen2023quantum}
Minh-Thi Nguyen, Jin-Guo Liu, Jonathan Wurtz, Mikhail~D Lukin, Sheng-Tao Wang, and Hannes Pichler.
\newblock Quantum optimization with arbitrary connectivity using rydberg atom arrays.
\newblock \emph{PRX Quantum}, 4\penalty0 (1):\penalty0 010316, 2023.
\newblock \doi{10.1103/prxquantum.4.010316}.

\bibitem[Lanthaler et~al.(2023)Lanthaler, Dlaska, Ender, and Lechner]{lanthaler2023rydberg}
Martin Lanthaler, Clemens Dlaska, Kilian Ender, and Wolfgang Lechner.
\newblock Rydberg-blockade-based parity quantum optimization.
\newblock \emph{Physical Review Letters}, 130\penalty0 (22):\penalty0 220601, 2023.
\newblock \doi{10.1103/physrevlett.130.220601}.

\bibitem[Byun et~al.(2024)Byun, Jung, Kim, Kim, Jeong, Jeong, and Ahn]{byun2024rydberg}
Andrew Byun, Junwoo Jung, Kangheun Kim, Minhyuk Kim, Seokho Jeong, Heejeong Jeong, and Jaewook Ahn.
\newblock Rydberg-atom graphs for quadratic unconstrained binary optimization problems.
\newblock \emph{Advanced Quantum Technologies}, 7\penalty0 (8):\penalty0 2300398, 2024.
\newblock \doi{10.1002/qute.202300398}.

\bibitem[Jeong et~al.(2023)Jeong, Kim, Hhan, Park, and Ahn]{jeong2023quantum}
Seokho Jeong, Minhyuk Kim, Minki Hhan, JuYoung Park, and Jaewook Ahn.
\newblock Quantum programming of the satisfiability problem with rydberg atom graphs.
\newblock \emph{Physical Review Research}, 5\penalty0 (4):\penalty0 043037, 2023.
\newblock \doi{10.1103/physrevresearch.5.043037}.

\bibitem[Finnila et~al.(1994)Finnila, Gomez, Sebenik, Stenson, and Doll]{finnila1994quantum}
Aleta~Berk Finnila, Maria~A Gomez, C~Sebenik, Catherine Stenson, and Jimmie~D Doll.
\newblock Quantum annealing: A new method for minimizing multidimensional functions.
\newblock \emph{Chemical physics letters}, 219\penalty0 (5-6):\penalty0 343--348, 1994.

\bibitem[Kadowaki and Nishimori(1998)]{kadowaki1998quantum}
Tadashi Kadowaki and Hidetoshi Nishimori.
\newblock Quantum annealing in the transverse ising model.
\newblock \emph{Physical Review E}, 58\penalty0 (5):\penalty0 5355, 1998.

\bibitem[Farhi et~al.(2000)Farhi, Goldstone, Gutmann, and Sipser]{farhi2000quantum}
Edward Farhi, Jeffrey Goldstone, Sam Gutmann, and Michael Sipser.
\newblock Quantum computation by adiabatic evolution.
\newblock \emph{arXiv preprint quant-ph/0001106}, 2000.

\bibitem[Santoro et~al.(2002)Santoro, Marton{\'a}k, Tosatti, and Car]{santoro2002theory}
Giuseppe~E Santoro, Roman Marton{\'a}k, Erio Tosatti, and Roberto Car.
\newblock Theory of quantum annealing of an ising spin glass.
\newblock \emph{Science}, 295\penalty0 (5564):\penalty0 2427--2430, 2002.

\bibitem[Farhi et~al.(2014{\natexlab{b}})Farhi, Goldstone, and Gutmann]{QAOA}
Edward Farhi, Jeffrey Goldstone, and Sam Gutmann.
\newblock A quantum approximate optimization algorithm.
\newblock \emph{ArXiv:1411.4028}, art. arXiv:1411.4028, 11 2014{\natexlab{b}}.
\newblock \doi{10.48550/ARXIV.1411.4028}.

\bibitem[Blekos et~al.(2024)Blekos, Brand, Ceschini, Chou, Li, Pandya, and Summer]{Blekos_2024}
Kostas Blekos, Dean Brand, Andrea Ceschini, Chiao-Hui Chou, Rui-Hao Li, Komal Pandya, and Alessandro Summer.
\newblock A review on quantum approximate optimization algorithm and its variants.
\newblock \emph{Physics Reports}, 1068:\penalty0 1–66, 6 2024.
\newblock ISSN 0370-1573.
\newblock \doi{10.1016/j.physrep.2024.03.002}.

\bibitem[Zhou et~al.(2020)Zhou, Wang, Choi, Pichler, and Lukin]{Zhou20}
Leo Zhou, Sheng-Tao Wang, Soonwon Choi, Hannes Pichler, and Mikhail~D. Lukin.
\newblock Quantum approximate optimization algorithm: Performance, mechanism, and implementation on near-term devices.
\newblock \emph{Phys. Rev. X}, 10:\penalty0 021067, 6 2020.
\newblock \doi{10.1103/PhysRevX.10.021067}.

\bibitem[Ebadi et~al.(2022)Ebadi, Keesling, Cain, Wang, Levine, Bluvstein, Semeghini, Omran, Liu, Samajdar, et~al.]{ebadi2022quantum}
Sepehr Ebadi, Alexander Keesling, Madelyn Cain, Tout~T Wang, Harry Levine, Dolev Bluvstein, Giulia Semeghini, Ahmed Omran, J-G Liu, Rhine Samajdar, et~al.
\newblock Quantum optimization of maximum independent set using rydberg atom arrays.
\newblock \emph{Science}, 376\penalty0 (6598):\penalty0 1209--1215, 2022.
\newblock \doi{10.1126/science.abo6587}.

\bibitem[Schiffer et~al.(2024)Schiffer, Wild, Maskara, Cain, Lukin, and Samajdar]{schiffer2024circumventing}
Benjamin~F Schiffer, Dominik~S Wild, Nishad Maskara, Madelyn Cain, Mikhail~D Lukin, and Rhine Samajdar.
\newblock Circumventing superexponential runtimes for hard instances of quantum adiabatic optimization.
\newblock \emph{Physical Review Research}, 6\penalty0 (1):\penalty0 013271, 2024.
\newblock \doi{10.1103/physrevresearch.6.013271}.

\bibitem[Andrist et~al.(2023)Andrist, Schuetz, Minssen, Yalovetzky, Chakrabarti, Herman, Kumar, Salton, Shaydulin, Sun, et~al.]{andrist2023hardness}
Ruben~S Andrist, Martin~JA Schuetz, Pierre Minssen, Romina Yalovetzky, Shouvanik Chakrabarti, Dylan Herman, Niraj Kumar, Grant Salton, Ruslan Shaydulin, Yue Sun, et~al.
\newblock Hardness of the maximum-independent-set problem on unit-disk graphs and prospects for quantum speedups.
\newblock \emph{Physical Review Research}, 5\penalty0 (4):\penalty0 043277, 2023.
\newblock \doi{10.1103/physrevresearch.5.043277}.

\bibitem[Selby(2014)]{selby2014efficient}
Alex Selby.
\newblock Efficient subgraph-based sampling of ising-type models with frustration.
\newblock \emph{arXiv preprint arXiv:1409.3934}, 2014.
\newblock \doi{10.1103/physrevb.63.224401}.

\bibitem[Mandra et~al.(2016)Mandra, Zhu, Wang, Perdomo-Ortiz, and Katzgraber]{mandra2016strengths}
Salvatore Mandra, Zheng Zhu, Wenlong Wang, Alejandro Perdomo-Ortiz, and Helmut~G Katzgraber.
\newblock Strengths and weaknesses of weak-strong cluster problems: A detailed overview of state-of-the-art classical heuristics versus quantum approaches.
\newblock \emph{Physical Review A}, 94\penalty0 (2):\penalty0 022337, 2016.
\newblock \doi{10.1103/physreva.94.022337}.

\bibitem[Albash and Lidar(2018)]{albash2018demonstration}
Tameem Albash and Daniel~A Lidar.
\newblock Demonstration of a scaling advantage for a quantum annealer over simulated annealing.
\newblock \emph{Physical Review X}, 8\penalty0 (3):\penalty0 031016, 2018.
\newblock \doi{10.1103/physrevx.8.031016}.

\bibitem[R{\o}nnow et~al.(2014)R{\o}nnow, Wang, Job, Boixo, Isakov, Wecker, Martinis, Lidar, and Troyer]{ronnow2014defining}
Troels~F R{\o}nnow, Zhihui Wang, Joshua Job, Sergio Boixo, Sergei~V Isakov, David Wecker, John~M Martinis, Daniel~A Lidar, and Matthias Troyer.
\newblock Defining and detecting quantum speedup.
\newblock \emph{science}, 345\penalty0 (6195):\penalty0 420--424, 2014.
\newblock \doi{10.1126/science.1252319}.

\bibitem[Dalzell et~al.(2020)Dalzell, Harrow, Koh, and La~Placa]{dalzell2020many}
Alexander~M Dalzell, Aram~W Harrow, Dax~Enshan Koh, and Rolando~L La~Placa.
\newblock How many qubits are needed for quantum computational supremacy?
\newblock \emph{Quantum}, 4:\penalty0 264, 2020.
\newblock \doi{10.22331/q-2020-05-11-264}.

\bibitem[Katzgraber et~al.(2015)Katzgraber, Hamze, Zhu, Ochoa, and Munoz-Bauza]{katzgraber2015seeking}
Helmut~G Katzgraber, Firas Hamze, Zheng Zhu, Andrew~J Ochoa, and Humberto Munoz-Bauza.
\newblock Seeking quantum speedup through spin glasses: The good, the bad, and the ugly.
\newblock \emph{Physical Review X}, 5\penalty0 (3):\penalty0 031026, 2015.
\newblock \doi{10.1103/physrevx.5.031026}.

\bibitem[Hen et~al.(2015)Hen, Job, Albash, R{\o}nnow, Troyer, and Lidar]{hen2015probing}
Itay Hen, Joshua Job, Tameem Albash, Troels~F R{\o}nnow, Matthias Troyer, and Daniel~A Lidar.
\newblock Probing for quantum speedup in spin-glass problems with planted solutions.
\newblock \emph{Physical Review A}, 92\penalty0 (4):\penalty0 042325, 2015.
\newblock \doi{10.1103/physreva.92.042325}.

\bibitem[Zhu et~al.(2016)Zhu, Ochoa, Schnabel, Hamze, and Katzgraber]{zhu2016best}
Zheng Zhu, Andrew~J Ochoa, Stefan Schnabel, Firas Hamze, and Helmut~G Katzgraber.
\newblock Best-case performance of quantum annealers on native spin-glass benchmarks: How chaos can affect success probabilities.
\newblock \emph{Physical Review A}, 93\penalty0 (1):\penalty0 012317, 2016.
\newblock \doi{10.1103/physreva.93.012317}.

\bibitem[CPLEX(2024)]{cplex2024}
IBM~ILOG CPLEX.
\newblock \emph{V24.1: User's Manual for CPLEX}.
\newblock International Business Machines Corporation, 2024.
\newblock URL \url{https://www.ibm.com/products/ilog-cplex-optimization-studio}.

\bibitem[Bhattacharya et~al.(2014)Bhattacharya, De, Nandy, and Roy]{bhattacharya2014maximum}
Binay~K Bhattacharya, Minati De, Subhas~C Nandy, and Sasanka Roy.
\newblock Maximum independent set for interval graphs and trees in space efficient models.
\newblock In \emph{CCCG}, 2014.
\newblock \doi{10.2139/ssrn.1968428}.

\bibitem[Garey and Johnson(1979)]{garey1979computers}
Michael~R Garey and David~S Johnson.
\newblock \emph{Computers and intractability}, volume 174.
\newblock freeman San Francisco, 1979.
\newblock \doi{10.1007/978-1-4612-0515-9_1}.

\bibitem[M{\'e}zard and Parisi(1986)]{mezard1986replica}
Marc M{\'e}zard and Giorgio Parisi.
\newblock A replica analysis of the travelling salesman problem.
\newblock \emph{Journal de physique}, 47\penalty0 (8):\penalty0 1285--1296, 1986.
\newblock \doi{10.1051/jphys:019860047080128500}.

\bibitem[Fu and Anderson(1986)]{fu1986application}
Yaotian Fu and Philip~W Anderson.
\newblock Application of statistical mechanics to np-complete problems in combinatorial optimisation.
\newblock \emph{Journal of Physics A: Mathematical and General}, 19\penalty0 (9):\penalty0 1605, 1986.
\newblock \doi{10.1142/9789812799371_0037}.

\bibitem[Monasson et~al.(1999)Monasson, Zecchina, Kirkpatrick, Selman, and Troyansky]{monasson1999determining}
R{\'e}mi Monasson, Riccardo Zecchina, Scott Kirkpatrick, Bart Selman, and Lidror Troyansky.
\newblock Determining computational complexity from characteristic ‘phase transitions’.
\newblock \emph{Nature}, 400\penalty0 (6740):\penalty0 133--137, 1999.
\newblock \doi{10.1038/22055}.

\bibitem[Krzaka{\l}a et~al.(2007)Krzaka{\l}a, Montanari, Ricci-Tersenghi, Semerjian, and Zdeborov{\'a}]{krzakala2007gibbs}
Florent Krzaka{\l}a, Andrea Montanari, Federico Ricci-Tersenghi, Guilhem Semerjian, and Lenka Zdeborov{\'a}.
\newblock Gibbs states and the set of solutions of random constraint satisfaction problems.
\newblock \emph{Proceedings of the National Academy of Sciences}, 104\penalty0 (25):\penalty0 10318--10323, 2007.
\newblock \doi{10.1073/pnas.0703685104}.

\bibitem[Achlioptas and Coja-Oghlan(2008)]{achlioptas2008algorithmic}
Dimitris Achlioptas and Amin Coja-Oghlan.
\newblock Algorithmic barriers from phase transitions.
\newblock In \emph{2008 49th Annual IEEE Symposium on Foundations of Computer Science}, pages 793--802. IEEE, 2008.
\newblock \doi{10.1109/focs.2008.11}.

\bibitem[Coja-Oghlan and Efthymiou(2015)]{coja2015independent}
Amin Coja-Oghlan and Charilaos Efthymiou.
\newblock On independent sets in random graphs.
\newblock \emph{Random Structures \& Algorithms}, 47\penalty0 (3):\penalty0 436--486, 2015.
\newblock \doi{10.1007/978-0-387-30162-4_187}.

\bibitem[Hartmann and Weigt(2006)]{hartmann2006phase}
Alexander~K Hartmann and Martin Weigt.
\newblock \emph{Phase transitions in combinatorial optimization problems: basics, algorithms and statistical mechanics}.
\newblock John Wiley \& Sons, 2006.
\newblock \doi{10.1016/s0304-3975(01)00149-9}.

\bibitem[Gent and Walsh(1994)]{gent1994sat}
Ian~P Gent and Toby Walsh.
\newblock The sat phase transition.
\newblock In \emph{ECAI}, volume~94, pages 105--109. PITMAN, 1994.
\newblock \doi{10.1007/bf02917402}.

\bibitem[Culberson and Gent(2001)]{culberson2001frozen}
Joseph Culberson and Ian Gent.
\newblock Frozen development in graph coloring.
\newblock \emph{Theoretical computer science}, 265\penalty0 (1-2):\penalty0 227--264, 2001.
\newblock \doi{10.1016/s0304-3975(01)00164-5}.

\bibitem[Mammen and Hogg(1997)]{mammen1997new}
Dorothy~L Mammen and Tad Hogg.
\newblock A new look at the easy-hard-easy pattern of combinatorial search difficulty.
\newblock \emph{Journal of Artificial Intelligence Research}, 7:\penalty0 47--66, 1997.
\newblock \doi{10.1613/jair.370}.

\bibitem[Vlasie(1995)]{vlasie1995systematic}
Romulus~Dan Vlasie.
\newblock Systematic generation of very hard cases for graph 3-colorability.
\newblock In \emph{Proceedings of 7th IEEE International Conference on Tools with Artificial Intelligence}, pages 114--119. IEEE, 1995.
\newblock \doi{10.1109/tai.1995.479412}.

\bibitem[Vlasie(1996)]{vlasie1996very}
Dan~R Vlasie.
\newblock The very particular structure of the very hard instances.
\newblock In \emph{Proceedings of the thirteenth national conference on Artificial intelligence-Volume 1}, pages 266--270, 1996.
\newblock \doi{10.1016/j.dam.2006.07.015}.

\bibitem[Bang-Jensen et~al.(2019)Bang-Jensen, Bellitto, Stiebitz, and Schweser]{bang2019haj}
J{\o}rgen Bang-Jensen, Thomas Bellitto, Michael Stiebitz, and Thomas Schweser.
\newblock Haj{\'{o}}s and ore constructions for digraphs.
\newblock \emph{arXiv preprint arXiv:1908.04096}, 2019.
\newblock \doi{10.37236/8942}.

\bibitem[Leo(2013)]{leo2013solving}
Gianmaria Leo.
\newblock \emph{Solving hard instances of maximum stable set problem by equitable partitions}.
\newblock PhD thesis, Universit{\`a} degli Studi di Roma" La Sapienza", 2013.

\bibitem[Sloane()]{sloane_graphs}
Neil J.~A. Sloane.
\newblock Challenge problems: Independent sets in graphs.
\newblock URL \url{https://oeis.org/A265032/a265032.html}.
\newblock Accessed: 2024-07-26.

\bibitem[{DIMACS}(1992)]{dimacs1992}
{DIMACS}.
\newblock Clique benchmark instances (web site), 1992.
\newblock URL \url{http://cs.hbg.psu.edu/txn131/clique.html}.
\newblock Accessed: 2024-07-26.

\bibitem[Marshall et~al.(2016)Marshall, Martin-Mayor, and Hen]{marshall2016practical}
Jeffrey Marshall, Victor Martin-Mayor, and Itay Hen.
\newblock Practical engineering of hard spin-glass instances.
\newblock \emph{Physical review A}, 94\penalty0 (1):\penalty0 012320, 2016.
\newblock \doi{10.1103/physreva.94.012320}.

\bibitem[Somma et~al.(2012)Somma, Nagaj, and Kieferov{\'a}]{somma2012quantum}
Rolando~D Somma, Daniel Nagaj, and M{\'a}ria Kieferov{\'a}.
\newblock Quantum speedup by quantum annealing.
\newblock \emph{Physical review letters}, 109\penalty0 (5):\penalty0 050501, 2012.
\newblock \doi{10.1103/physrevlett.109.050501}.

\bibitem[Pang et~al.(2021)Pang, Coffrin, Lokhov, and Vuffray]{pang2021potential}
Yuchen Pang, Carleton Coffrin, Andrey~Y Lokhov, and Marc Vuffray.
\newblock The potential of quantum annealing for rapid solution structure identification.
\newblock \emph{Constraints}, 26\penalty0 (1):\penalty0 1--25, 2021.
\newblock \doi{10.1007/s10601-020-09315-0}.

\bibitem[Berkley et~al.(2010)Berkley, Johnson, Bunyk, Harris, Johansson, Lanting, Ladizinsky, Tolkacheva, Amin, and Rose]{berkley2010scalable}
AJ~Berkley, MW~Johnson, P~Bunyk, R~Harris, J~Johansson, T~Lanting, E~Ladizinsky, E~Tolkacheva, MHS Amin, and G~Rose.
\newblock A scalable readout system for a superconducting adiabatic quantum optimization system.
\newblock \emph{Superconductor Science and Technology}, 23\penalty0 (10):\penalty0 105014, 2010.
\newblock \doi{10.1088/0953-2048/23/10/105014}.

\bibitem[van Leeuwen(2005)]{van2005approximation}
Erik~Jan van Leeuwen.
\newblock Approximation algorithms for unit disk graphs.
\newblock In \emph{International Workshop on Graph-Theoretic Concepts in Computer Science}, pages 351--361. Springer, 2005.
\newblock \doi{10.1007/11604686_31}.

\bibitem[Arnborg and Proskurowski(1989)]{arnborg1989linear}
Stefan Arnborg and Andrzej Proskurowski.
\newblock Linear time algorithms for np-hard problems restricted to partial k-trees.
\newblock \emph{Discrete applied mathematics}, 23\penalty0 (1):\penalty0 11--24, 1989.
\newblock \doi{10.1016/0166-218x(89)90031-0}.

\bibitem[Lipton and Tarjan(1977)]{lipton1977applications}
Richard~J Lipton and Robert~Endre Tarjan.
\newblock Applications of a planar separator theorem.
\newblock In \emph{18th Annual Symposium on Foundations of Computer Science (sfcs 1977)}, pages 162--170. IEEE, 1977.
\newblock \doi{10.1137/0209046}.

\bibitem[de~Berg et~al.(2018)de~Berg, Bodlaender, Kisfaludi-Bak, Marx, and Zanden]{de2018framework}
Mark de~Berg, Hans~L Bodlaender, S{\'a}ndor Kisfaludi-Bak, D{\'a}niel Marx, and Tom C van~der Zanden.
\newblock A framework for eth-tight algorithms and lower bounds in geometric intersection graphs.
\newblock In \emph{Proceedings of the 50th Annual ACM SIGACT Symposium on Theory of Computing}, pages 574--586, 2018.
\newblock \doi{10.1145/3188745.3188854}.

\bibitem[Note2()]{Note2}
Note2.
\newblock An object is said to be $\alpha $-fat if the ratio between its circumscribed radius and inscribed radius is at most $\alpha $. Intuitively, an $\alpha $-fat object cannot be excessively elongated or thin. A set of objects is said to be similarly-sized if the ratio between the diameters of the largest and smallest object in the set is bounded by some constant $\alpha > 0$.

\bibitem[Cygan et~al.(2015)Cygan, Fomin, Kowalik, Lokshtanov, Marx, Pilipczuk, Pilipczuk, and Saurabh]{cygan2015parameterized}
Marek Cygan, Fedor~V. Fomin, Lukasz Kowalik, Daniel Lokshtanov, D{\'{a}}niel Marx, Marcin Pilipczuk, Michal Pilipczuk, and Saket Saurabh.
\newblock \emph{Parameterized Algorithms}.
\newblock Springer, 2015.
\newblock ISBN 978-3-319-21274-6.
\newblock \doi{10.1007/978-3-319-21275-3}.
\newblock URL \url{https://doi.org/10.1007/978-3-319-21275-3}.

\bibitem[van Leeuwen(2004)]{th:slabDecompoPolyvan2004optimization}
E.J. van Leeuwen.
\newblock \emph{Optimization Problems on Mobile Ad Hoc Networks -- Algorithms for Disk Graphs}.
\newblock PhD thesis, Master’s Thesis INF/SCR-04-32, Inst. of Information and Computing Sciences, Utrecht Univ., 2004.

\bibitem[H{\aa}stad(1999)]{haastad1999clique}
Johan H{\aa}stad.
\newblock Clique is hard to approximate within n 1- $\\varepsilon$.
\newblock \emph{Acta Math.}, 182\penalty0 (1):\penalty0 105--142, 1999.
\newblock \doi{10.1007/bf02392825}.

\bibitem[Hochbaum and Maass(1985)]{hochbaum1985approximation}
Dorit~S Hochbaum and Wolfgang Maass.
\newblock Approximation schemes for covering and packing problems in image processing and vlsi.
\newblock \emph{Journal of the ACM (JACM)}, 32\penalty0 (1):\penalty0 130--136, 1985.
\newblock \doi{10.1145/2455.214106}.

\bibitem[Bonamy et~al.(2021)Bonamy, Bonnet, Bousquet, Charbit, Giannopoulos, Kim, Rz{\k{a}}{\.z}ewski, Sikora, and Thomass{\'e}]{th:weightedPTASbonamy2021eptas}
Marthe Bonamy, {\'E}douard Bonnet, Nicolas Bousquet, Pierre Charbit, Panos Giannopoulos, Eun~Jung Kim, Pawe{\l} Rz{\k{a}}{\.z}ewski, Florian Sikora, and St{\'e}phan Thomass{\'e}.
\newblock Eptas and subexponential algorithm for maximum clique on disk and unit ball graphs.
\newblock \emph{Journal of the ACM (JACM)}, 68\penalty0 (2):\penalty0 1--38, 2021.
\newblock \doi{10.1145/3433160}.

\bibitem[van Leeuwen et~al.(2009)]{th:geometricThesis}
Erik~Jan van Leeuwen et~al.
\newblock \emph{Optimization and approximation on systems of geometric objects}.
\newblock Universiteit van Amsterdam [Host], 2009.
\newblock \doi{10.1007/978-1-4471-0495-7_26}.

\bibitem[et~al.()]{Pierretoappear}
Pierre~Cazals et~al.
\newblock to appear.
\newblock \doi{10.7767/9783205201731}.

\bibitem[Wannier(1950)]{wannier1950antiferromagnetism}
GH0038 Wannier.
\newblock Antiferromagnetism. the triangular ising net.
\newblock \emph{Physical Review}, 79\penalty0 (2):\penalty0 357, 1950.

\bibitem[Hwang et~al.(2024)Hwang, Hwang, Ahn, Yoshida, and Burgdorfer]{hwang2024impact}
Hansub Hwang, Sunhwa Hwang, Jaewook Ahn, Shuhei Yoshida, and Joachim Burgdorfer.
\newblock Impact-parameter selective rydberg atom collision by optical tweezers.
\newblock \emph{arXiv preprint arXiv:2412.06225}, 2024.
\newblock \doi{10.1016/s0065-2539(08)60109-2}.

\bibitem[Hagberg et~al.(2008)Hagberg, Swart, and Schult]{hagberg2008exploring}
Aric Hagberg, Pieter~J Swart, and Daniel~A Schult.
\newblock Exploring network structure, dynamics, and function using networkx.
\newblock 2008.
\newblock \doi{10.25080/tcwv9851}.

\bibitem[Lawler and Wood(1966)]{lawler1966branch}
Eugene~L Lawler and David~E Wood.
\newblock Branch-and-bound methods: A survey.
\newblock \emph{Operations research}, 14\penalty0 (4):\penalty0 699--719, 1966.
\newblock \doi{10.1287/opre.14.4.699}.

\bibitem[Barredo et~al.(2018)Barredo, Lienhard, De~Leseleuc, Lahaye, and Browaeys]{barredo2018synthetic}
Daniel Barredo, Vincent Lienhard, Sylvain De~Leseleuc, Thierry Lahaye, and Antoine Browaeys.
\newblock Synthetic three-dimensional atomic structures assembled atom by atom.
\newblock \emph{Nature}, 561\penalty0 (7721):\penalty0 79--82, 2018.
\newblock \doi{10.1038/s41586-018-0450-2}.

\bibitem[Nogrette et~al.(2014)Nogrette, Labuhn, Ravets, Barredo, B\\'eguin, Vernier, Lahaye, and Browaeys]{Nogrette14}
F.~Nogrette, H.~Labuhn, S.~Ravets, D.~Barredo, L.~B\\'eguin, A.~Vernier, T.~Lahaye, and A.~Browaeys.
\newblock Single-atom trapping in holographic 2d arrays of microtraps with arbitrary geometries.
\newblock \emph{Phys. Rev. X}, 4:\penalty0 021034, 5 2014.
\newblock \doi{10.1103/PhysRevX.4.021034}.

\bibitem[Henriet et~al.(2020)Henriet, Beguin, Signoles, Lahaye, Browaeys, Reymond, and Jurczak]{henriet2020quantum}
Lo{\"\i}c Henriet, Lucas Beguin, Adrien Signoles, Thierry Lahaye, Antoine Browaeys, Georges-Olivier Reymond, and Christophe Jurczak.
\newblock Quantum computing with neutral atoms.
\newblock \emph{Quantum}, 4:\penalty0 327, 2020.
\newblock \doi{10.22331/q-2020-09-21-327}.

\bibitem[{Morgado} and {Whitlock}(2021)]{Morgado2021}
M.~{Morgado} and S.~{Whitlock}.
\newblock {Quantum simulation and computing with Rydberg-interacting qubits}.
\newblock \emph{AVS Quantum Science}, 3\penalty0 (2):\penalty0 023501, 6 2021.
\newblock \doi{10.1116/5.0036562}.

\bibitem[B{\'{e} }guin et~al.(2013)B{\'{e} }guin, Vernier, Chicireanu, Lahaye, and Browaeys]{Beguin_2013}
L.~B{\'{e} }guin, A.~Vernier, R.~Chicireanu, T.~Lahaye, and A.~Browaeys.
\newblock Direct measurement of the van der waals interaction between two rydberg atoms.
\newblock \emph{Phys. Rev. Lett.}, 110\penalty0 (26), 6 2013.
\newblock \doi{10.1103/physrevlett.110.263201}.

\bibitem[{\v{S}}ibali{\'{c}} et~al.(2017){\v{S}}ibali{\'{c}}, Pritchard, Adams, and Weatherill]{ibali2017}
N.~{\v{S}}ibali{\'{c}}, J.D. Pritchard, C.S. Adams, and K.J. Weatherill.
\newblock {ARC}: An open-source library for calculating properties of alkali rydberg atoms.
\newblock \emph{Computer Physics Communications}, 220:\penalty0 319--331, 11 2017.
\newblock \doi{10.1016/j.cpc.2017.06.015}.

\bibitem[Quelle et~al.(2024)Quelle, Hénaff, Mendizabal, Grava, Merhej, Mendizábal, Wennersteen, and Teller]{emu-mps}
Anton Quelle, Pablo~Le Hénaff, Mauro Mendizabal, Stefano Grava, Elie Merhej, Mauro Mendizábal, Aleksander Wennersteen, and David Teller.
\newblock pasqal-io/emulators: v1.2.4, 2024.

\bibitem[Young et~al.(2020)Young, Eckner, Milner, Kedar, Norcia, Oelker, Schine, Ye, and Kaufman]{young2020half}
Aaron~W Young, William~J Eckner, William~R Milner, Dhruv Kedar, Matthew~A Norcia, Eric Oelker, Nathan Schine, Jun Ye, and Adam~M Kaufman.
\newblock Half-minute-scale atomic coherence and high relative stability in a tweezer clock.
\newblock \emph{Nature}, 588\penalty0 (7838):\penalty0 408--413, 2020.
\newblock \doi{https://doi.org/10.1038/s41586-020-3009-y}.

\bibitem[Gyger et~al.(2024)Gyger, Ammenwerth, Tao, Timme, et~al.]{gyger2024continuous}
Flavien Gyger, Maximilian Ammenwerth, Renhao Tao, Hendrik Timme, et~al.
\newblock Continuous operation of large-scale atom arrays in optical lattices, 2024.

\bibitem[Cimring et~al.(2023)Cimring, El~Sabeh, Bacvanski, Maaz, et~al.]{Cimring_2023}
Barry Cimring, Remy El~Sabeh, Marc Bacvanski, Stephanie Maaz, et~al.
\newblock Efficient algorithms to solve atom reconfiguration problems. i. redistribution-reconfiguration algorithm.
\newblock \emph{Phys. Rev. A}, 108\penalty0 (2), 2023.
\newblock ISSN 2469-9934.
\newblock \doi{10.1103/physreva.108.023107}.

\bibitem[Bergschneider et~al.(2018)Bergschneider, Klinkhamer, Becher, Klemt, et~al.]{Bergschneider_2018}
Andrea Bergschneider, Vincent~M. Klinkhamer, Jan~Hendrik Becher, Ralf Klemt, et~al.
\newblock Spin-resolved single-atom imaging of 6li in free space.
\newblock \emph{Phys. Rev. A}, 97\penalty0 (6), 2018.
\newblock ISSN 2469-9934.
\newblock \doi{10.1103/physreva.97.063613}.

\bibitem[Buzulutskov(2012)]{Buzulutskov_2012}
A~Buzulutskov.
\newblock Advances in cryogenic avalanche detectors.
\newblock \emph{Journal of Instrumentation}, 7\penalty0 (02):\penalty0 C02025–C02025, 2012.
\newblock ISSN 1748-0221.
\newblock \doi{10.1088/1748-0221/7/02/c02025}.

\bibitem[Manetsch et~al.(2024)Manetsch, Nomura, Bataille, Leung, Lv, and Endres]{manetsch2024tweezer}
Hannah~J Manetsch, Gyohei Nomura, Elie Bataille, Kon~H Leung, Xudong Lv, and Manuel Endres.
\newblock A tweezer array with 6100 highly coherent atomic qubits.
\newblock \emph{arXiv preprint arXiv:2403.12021}, 2024.

\bibitem[Bauza and Lidar(2024)]{bauza2024scaling}
Humberto~Munoz Bauza and Daniel~A Lidar.
\newblock Scaling advantage in approximate optimization with quantum annealing.
\newblock \emph{arXiv preprint arXiv:2401.07184}, 2024.

\bibitem[Scholl et~al.(2022)Scholl, Williams, Bornet, Wallner, Barredo, Henriet, Signoles, Hainaut, Franz, Geier, et~al.]{scholl2022microwave}
Pascal Scholl, Hannah~J Williams, Guillaume Bornet, Florian Wallner, Daniel Barredo, L~Henriet, Adrien Signoles, Cl{\'e}ment Hainaut, Titus Franz, S~Geier, et~al.
\newblock Microwave engineering of programmable xxz hamiltonians in arrays of rydberg atoms.
\newblock \emph{PRX Quantum}, 3\penalty0 (2):\penalty0 020303, 2022.
\newblock \doi{10.1109/cleoe-eqec.2017.8087332}.

\bibitem[Lee et~al.(2016)Lee, Kim, and Ahn]{lee2016three}
Woojun Lee, Hyosub Kim, and Jaewook Ahn.
\newblock Three-dimensional rearrangement of single atoms using actively controlled optical microtraps.
\newblock \emph{Optics express}, 24\penalty0 (9):\penalty0 9816--9825, 2016.

\bibitem[Bluvstein et~al.(2022)Bluvstein, Levine, Semeghini, Wang, Ebadi, Kalinowski, Keesling, Maskara, Pichler, Greiner, et~al.]{bluvstein2022quantum}
Dolev Bluvstein, Harry Levine, Giulia Semeghini, Tout~T Wang, Sepehr Ebadi, Marcin Kalinowski, Alexander Keesling, Nishad Maskara, Hannes Pichler, Markus Greiner, et~al.
\newblock A quantum processor based on coherent transport of entangled atom arrays.
\newblock \emph{Nature}, 604\penalty0 (7906):\penalty0 451--456, 2022.

\bibitem[Maciejewski et~al.(2020)Maciejewski, Zimbor{\'a}s, and Oszmaniec]{maciejewski2020mitigation}
Filip~B Maciejewski, Zolt{\'a}n Zimbor{\'a}s, and Micha{\l} Oszmaniec.
\newblock Mitigation of readout noise in near-term quantum devices by classical post-processing based on detector tomography.
\newblock \emph{Quantum}, 4:\penalty0 257, 2020.
\newblock \doi{10.22331/q-2020-04-24-257}.
\newblock URL \url{https://quantum-journal.org/papers/q-2020-04-24-257/}.

\bibitem[Willis(2011)]{willis2011bounds}
William Willis.
\newblock Bounds for the independence number of a graph.
\newblock Master thesis, Virginia Commonwealth University, 2011.

\bibitem[Balakrishnan et~al.(2004)Balakrishnan, Barrett, Kumar, Marathe, and Thite]{balakrishnan2004distance}
Hari Balakrishnan, Christopher~L Barrett, VS~Anil Kumar, Madhav~V Marathe, and Shripad Thite.
\newblock The distance-2 matching problem and its relationship to the mac-layer capacity of ad hoc wireless networks.
\newblock \emph{IEEE Journal on Selected Areas in Communications}, 22\penalty0 (6):\penalty0 1069--1079, 2004.
\newblock \doi{10.1109/jsac.2004.830909}.

\bibitem[Levit and Mandrescu(2018)]{levit2018annihilation}
Vadim~E Levit and Eugen Mandrescu.
\newblock On an annihilation number conjecture.
\newblock \emph{arXiv preprint arXiv:1811.04722}, 2018.
\newblock \doi{10.26493/1855-3974.1950.8bd}.

\bibitem[Larson and Pepper(2011)]{larson2011graphs}
Craig~E Larson and Ryan Pepper.
\newblock Graphs with equal independence and annihilation numbers.
\newblock \emph{the electronic journal of combinatorics}, pages P180--P180, 2011.
\newblock \doi{10.37236/667}.

\bibitem[Singh and Pandey(2015)]{singh2015survey}
Krishna~Kumar Singh and Ajeet~Kumar Pandey.
\newblock Survey of algorithms on maximum clique problem.
\newblock \emph{International Advanced Research Journal in Science, Engineering and Technology}, 2\penalty0 (2):\penalty0 18--20, 2015.
\newblock \doi{10.17148/iarjset.2015.2203}.

\end{thebibliography}

\onecolumngrid
\appendix
\renewcommand{\thefigure}{A\arabic{figure}}
\setcounter{figure}{0} 
\newpage
\section{Maximum Independent Set problem formulation}
\label{sec:prob_statement}
Let us introduce the Maximum Weighted Independent Set (MWIS) problem and its optimization form as a quadratic unconstrained binary optimization (QUBO) problem.
Let \( G = (V, E) \) be an undirected graph, where \( V \) is the set of vertices and \( E \subseteq V \times V \) is the set of edges. An independent set is a subset \( S \subseteq V \) such that no two vertices in \( S \) are adjacent:
\[
(i, j) \in E \implies i \notin S \text{ or } j \notin S.
\]
The MWIS problem consists in finding the largest independent set, which can be formulated as:
\begin{equation}
\begin{aligned}
& \max_{x \in \{0,1\}^{|V|}} \sum_{i \in V} w_i x_i, \quad \text{subject to}\\
& x_i + x_j \leq 1, \quad \forall (i, j) \in E,
\end{aligned}
\end{equation}
where $w_i \in [0,1]$ and \( x_i \) are binary decision variables defined as:
\[
x_i =
\begin{cases}
1, & \text{if } i \in S, \\
0, & \text{otherwise}.
\end{cases}
\]
The QUBO model of the MWIS implies introducing a penalty term for violating the independence constraint:
\begin{align}
\max_{x \in \{0,1\}^{|V|}} & \sum_{i \in V} w_i x_i - \alpha \sum_{(i,j) \in E} x_i x_j,
\end{align}
where \( \alpha > 0 \) is a penalty parameter.
Since QUBO problems are typically expressed in a minimization form, an equivalent formulation is:
\begin{equation}
\label{eq:mis_class}
\min_{x \in \{0,1\}^{|V|}} \alpha  \sum_{(i,j) \in E} x_i x_j - \sum_{i \in V} w_i x_i.
\end{equation}
Here, we denote with $x^{*}$ the optimal solution of the problem. Let us observe that, in the case of the weights $w_i$ being uniformly $1$, the problem to solve becomes the above-mentioned MIS.

\section{The impact of density on non-embeddable instances}
\label{app:density}
\begin{figure}[ht]
    \centering
    \includegraphics[width=0.5\linewidth]{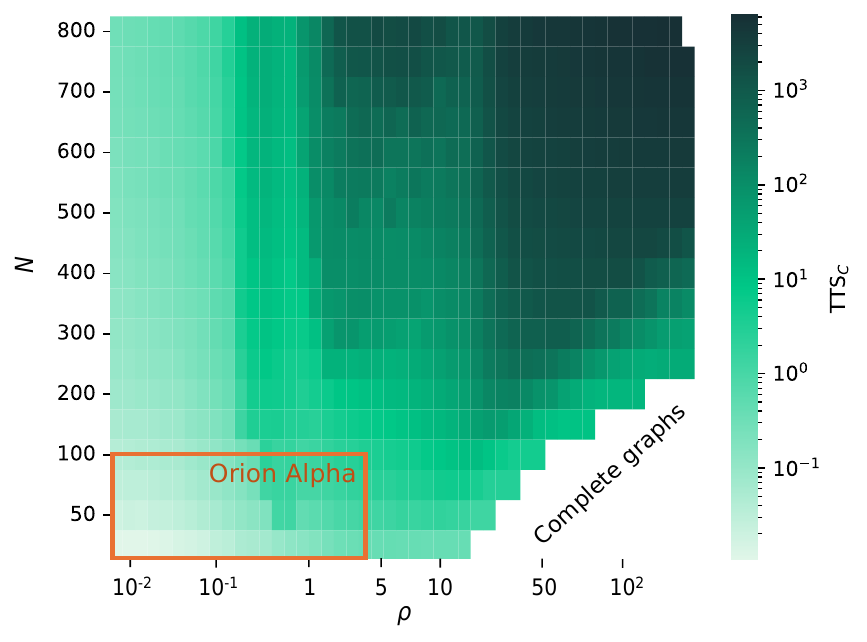}
    \caption{Density as a hardness parameter for MIS in UD graphs.   
For each pair of size \( N \) and density \( \rho \), the classical solver is tasked with finding the MIS of $100$ randomly generated graphs. The color map depicts the TTS\(_C\), where light green indicates fast resolution and dark green denotes more challenging instances. The white region corresponds to size and density values that produce the complete graph where nodes are all-to-all connected, for which the MIS problem is trivial to solve.
 }
    \label{fig:density_runtime}
\end{figure}

We verify the theoretical insights by running extensive numerical benchmarks,  considering MIS on random UD graphs, constructed by placing $n$ vertices randomly with a density $\rho$ in a 2D box of size $L^2,$ where $L =
\sqrt{n/\rho} $ and the UD radius is fixed to $r = 1$, similarly to the construction proposed in Ref.~\cite{pichler2018quantum}.
The result is shown in Fig.\,\ref{fig:density_runtime}, where we discover the existence of a hardness phase in the size/density landscape. The numerical benchmark we ran is the most important one done compared to existing literature, with $100$ graphs per point in the plot and going up to size $n=1000$. The density is varied from sparse to complete graphs. If the density is below the percolation threshold $\rho_c \approx 1.43$, the graph decomposes into disconnected finite clusters, allowing for a fast solve. A phase of harder graphs is probed with the darker zone. With the current hardware capabilities of Orion Alpha, the accessible graph space is limited to relatively easy-to-solve instances. However, pushing the number of atoms and/or the density achievable on future generations will enable the exploration of significantly more challenging graph instances.

\section{Transition to Erdos-Renyi is correlated with higher treewidth in graphs}
\label{app:UD_to_ER}

Inspired by the works from Ref~\cite{andrist2023hardness}, we show that an increase in treewidth is associated with an increase in runtime. These instances are however not realisable on neutral atom platforms as such.

A way of doing so is to follow a method proposed in Ref.\,\cite{andrist2023hardness}: starting from KL graphs, edges are randomly selected and rewired, thereby gradually
breaking the UD connectivity and ultimately generating random ER graphs. The advantage of this method is that the density of edges in the graph is unchanged, but the locality of the UD graphs is broken. It also transitions from graphs that are easy to solve to graphs that are hard to solve. 

\begin{figure*}[!]
     \centering
     \includegraphics[width=0.8\linewidth]{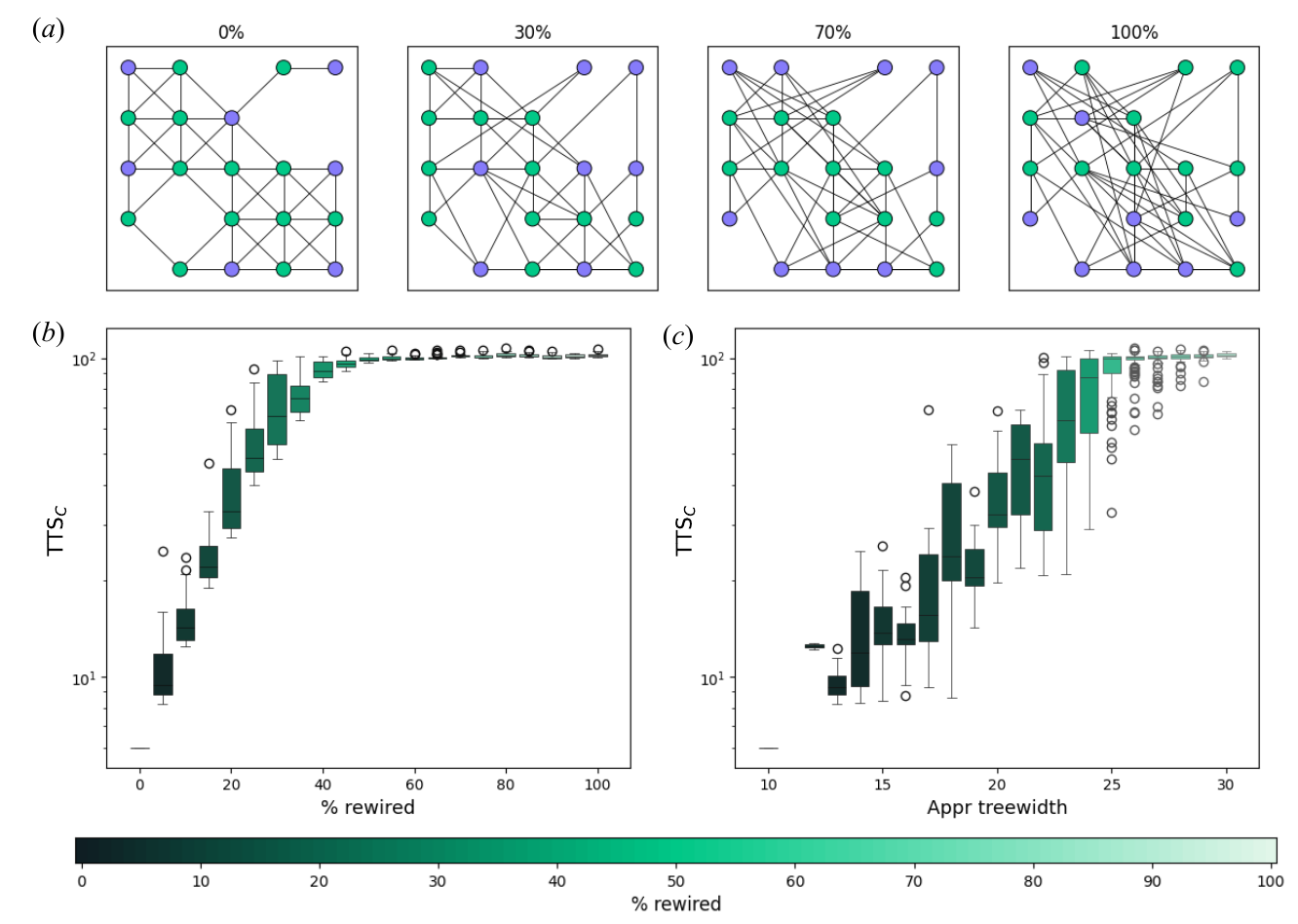}
     \caption{Evolution of the treewidth when transitioning from a KL to an ER instance.
        In this procedure, we re-wire the edges randomly without modifying the graph density. 
        By re-wiring the edges, the graph loses the initial locality in its topology, making it harder to solve. 
        As such, we observe that stronger density or non-locality increase the treewidth of the graph.  
        $(a)$ Graphs at different re-wiring percentages ($0\%$, $30\%$, $70\%$, $100\%$) show the change in structure as edges are randomly re-wired while maintaining graph density. 
        $(b)$ The TTS$_c$ (in ticks) as a function of the percentage of re-wired edges indicates that higher re-wiring percentages make the graph harder to solve. Each box plots show the distribution of TTS$_C$ across $100$ graphs with eventual outliers (with surprisingly high TTS$_C$) discarded from the boxes and shown as white circles.
        $(c)$ Similar plot for TTS$_C$ as a function of approximate treewidth.
    }
     \label{fig:treewidth}
\end{figure*}

The increase of runtime in Fig.\,\ref{fig:treewidth} as the percentage of re-wired edges increases is a numerical confirmation of this intuition. We discover in this benchmark that the treewidth also increases as the rewiring increases, and we therefore find a good correlation between treewidth and runtime. We believe that increasing the treewidth will be a good way to increase the difficulty of solving the MIS problem. The caveat is that it is NP-hard to calculate the treewidth, but it can be brute-forced up to graphs of size $N=350$. 
Furthermore, while treewidth remains a good indicator of difficulty known in parameterised complexity, it doesn't capture everything hard about a problem: for example, cliques have maximum treewidth, but most problems are easy to solve on them.

\section{Neutral atoms quantum platforms and quantum annealing}
\label{app:quantum}
In this section, we describe more technical details of the quantum simulations and data post-processing presented in the main text.

\subsection{Generation of the native instances}
\label{sec:generate_instances}
\begin{figure*}
    \includegraphics[width=0.7\linewidth]{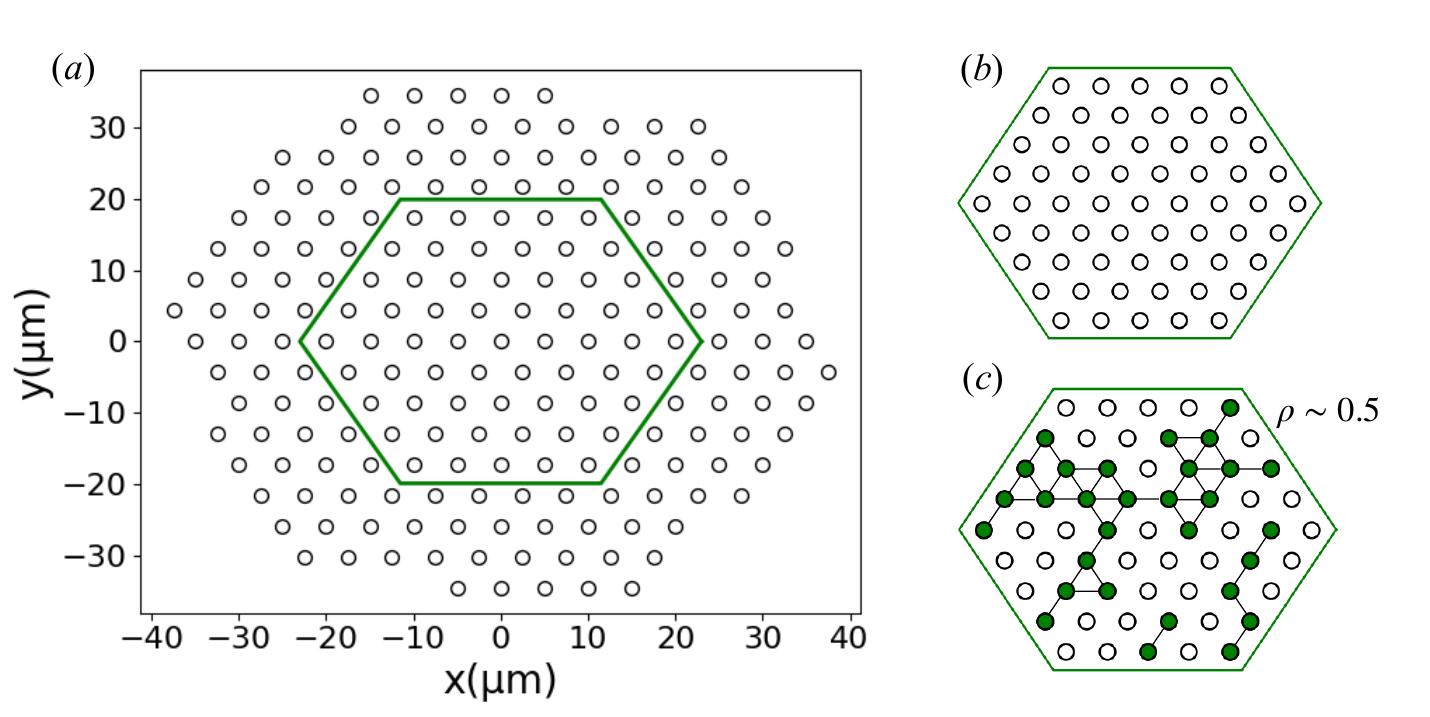}
    \caption{ Example of generation of an instance with $N=61$ nodes and density $\rho=\frac{30}{61}\sim 0.5$. $(a)$ We define the full register; $(b)$ we select the traps to be taken into account, then $(c)$ we sample $N$ indices out of the $L$ available and link them. As such, we obtain a UD graph that is natively embeddable on the quantum machine.
    }
    \label{fig:instancegenerator}
\end{figure*}

In Sec.~\ref{sec:experiments} we, so called, native instances of the neutral atoms quantum platforms with different system sizes $N$ and densities $\rho$ considered.
To this end, we start from a selected pre-calibrated layout of the quantum platform, corresponding to a triangular layout composed of 200 traps, spaced $5\mu$m from each other.
Out of these traps, we select the ones we want to work with at fixed $\rho$ and $N$, considering only the closest $L=\frac{N}{\rho}$ to the center. In this way, selecting $N$ traps out of $L$, we would be able to obtain a density of filled traps equal to $\rho$.
Finally, we sample $N$ trap indices out of $L$ to initialize the positions of the neutral atoms.
An example of this is shown in Fig.~\ref{fig:instancegenerator}.
We select the traps to be taken into account, highlighted by the green hexagon. Then, we sample $N$ indices out of the $L$ available and link them with edges in black. The final graph is a UD graph that is natively embeddable on the quantum machine.
The resulting graphs are natively embeddable on the hardware by using
the layout in Fig.A3(a) as the trap layout and the interactions between atoms trapped at the selected sites exhibit the
same adjacency matrix as the graph to reproduce (up to the nearest neighbour approximation performed by neglecting
terms of order 1/27). Thus, we arrange the item in such a pattern and perform a suitable evolution to drive the system towards the target quantum state. The latter is described in the following section.

\subsection{Annealing schedule}
\begin{figure}
    \centering
    \includegraphics[width=0.7\textwidth]{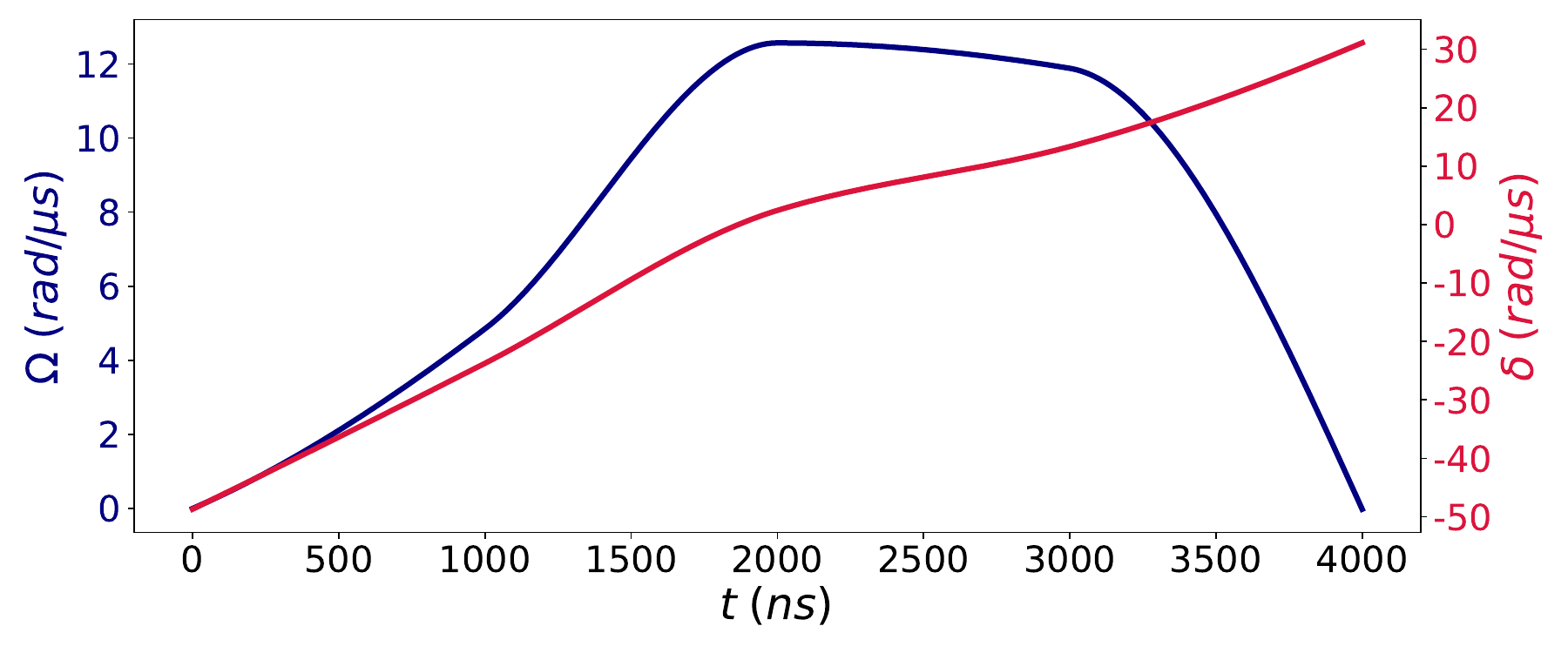}
    \caption{Annealing protocol of time dependent controls $\Omega$ and $\delta$ employed to find the MIS of the graphs.}
    \label{fig:annealing-protocol}
\end{figure}

As discussed in the main text, we employed quantum annealing for solving the MWIS problem.
It entails preparing an initial state with high fidelity and evolving it adiabatically towards the state that encodes the target solution.
In our case, this is obtained by changing the control-field in time.
The particular time-dependence is called annealing schedule.
In this work, we have adopted the one utilized shown in Fig.~\ref{fig:annealing-protocol}. It has been obtained by performing Bayesian optimization on the values of $\delta$ and $\Omega$ at the interpolation points marked in black.
The initial and final values of $\Omega$ have been fixed such that the initial Hamiltonian ground state corresponds to the prepared initial state, and the final Hamiltonian encodes the optimization problem.
The initial detuning value instead has been fixed to the minimum possible value, compatible with hardware capabilities.
All the other points have been optimized,to maximize the probability of finding the MIS, over a set of random unit-disk graphs, generated as described in Sec.~\ref{sec:generate_instances}, with $N=20$ and densities varying from 0.1 to 1 in steps of 0.1.
Arranging the atoms in native instances and running the optimised sequence according to the prescriptions in this section, allows preparing the quantum system in a state that encodes the solution of the MWIS problem.
Sampling from that quantum superposition returns a distribution of bitstrings which either constitute perfect (MWIS) or approximated (MWIS-$k$) solutions for the problem.

\section{Noise mitigation and post-processing}
\label{sec:nm_and_pp}
To the end of improving the results of the QPU, we mitigate the measurement errors and employ some post-processing routines to the samples bitstrings.

Regarding measurement errors.
following Ref.~\cite{maciejewski2020mitigation}, we assume the measurement errors in the device are perfectly characterized beforehand by such that a local correction matrix can be computed. Furthermore, we consider the noise channel to be completely classical.
Under these assumptions, we can straightforwardly derive a correction matrix 
\begin{equation}
    \Lambda^{-1}=\frac{1}{1-p-q}\begin{pmatrix}1-q & -q \\-p & 1-p 
    \end{pmatrix}
\end{equation}
and apply it
to the state vector the represents the probability distribution of our state. The former is obtained through the sampling of the desired state prepared on the QPU.
In our case, $p$ and $q$ represent:
the probability of detecting an atom in the ground state as if it were in the excited state (false positives) $p\sim 0.03$; and the probability of detecting an atom in the excited state as if it were in the ground state (“false negatives”) $q\sim 0.08$.
To scale this error mitigation approach to larger system sizes, we represent the state vector as an MPS and the noise channel as an MPO.

For what concerns post-processing routines, we apply the same approaches discussed in Ref.~\cite{leclerc2024implementingtransferableannealingprotocols}, there the authors
aim to correct the sampled bitstrings, by manually flipping bits, to the end of improving their quality in regard to the MIS problem. For each bitstring, we verify its validity as an independent set (IS). If necessary, we address any constraint violations by ensuring that adjacent elements are properly taken into account.

\begin{figure}
    \centering
    \includegraphics[width=0.5\linewidth]{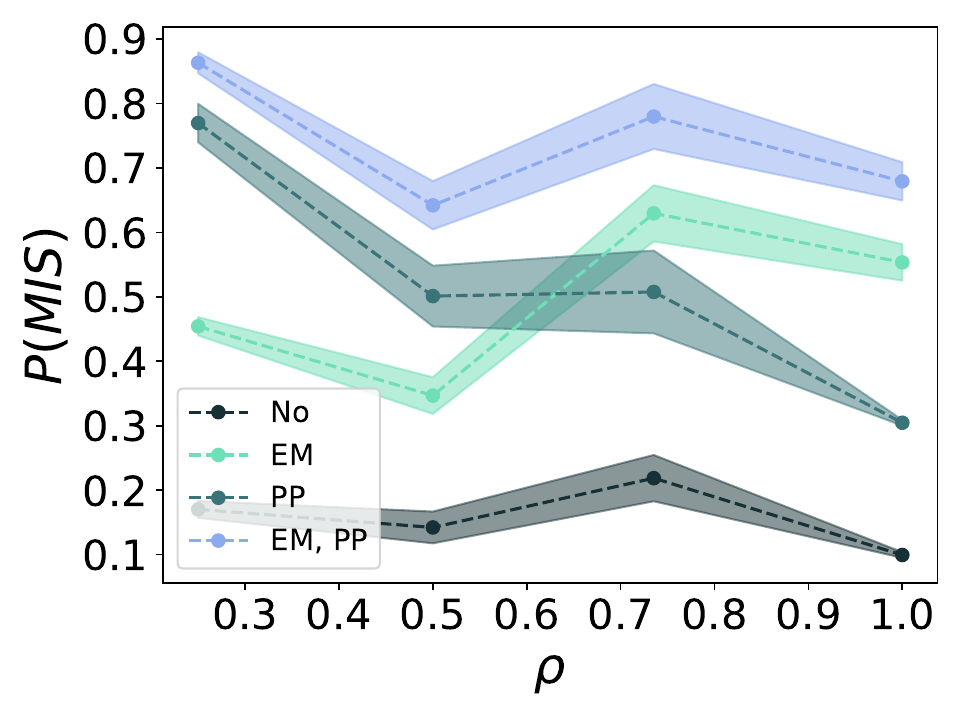}
    \caption{Probability of finding the MIS as a function of the density of the graph, for $N=25$ sites. We compare the original results ('No'), with measurement error mitigation ('EM'), post-processing ('PP'), and both ('EM, PP').}
    \label{fig:error_mitigation}
\end{figure}

In Fig.~\ref{fig:error_mitigation} we show an example of how the results change adding measurement errors mitigation and post-processing. In particular, we plot the probability of finding the MIS for 25 sites as a function of density, as defined in the previous section, for four different cases. 
We observe that implementing post-processing and error mitigation significantly enhances performance, tripling the probability of finding the maximum independent set (MIS) for smaller system sizes. This is less effective increasing the number of nodes, eventually becoming irrelevant for the largest sizes we show.

\section{Constructive methods for generating hard weighted MIS instances}
\label{appendix:weight_methods}

This appendix presents additional constructive methods for generating hard MWIS instances. The intuition behind these methods is derived from known upper bounds on the unweighted MIS problem. In \cite{willis2011bounds}, various upper bounds on the independence number (i.e., the optimal value of the unweighted MIS) of a graph are discussed, which aid branch-and-bound solvers by efficiently pruning the search space. Similarly, for the weighted version of the MIS, we aim to construct weight assignments that weaken the linear relaxation, resulting in a weaker upper bound. Additionally, we propose methods that introduce symmetries (yielding equivalent solutions), which may further increase the complexity of solving the MWIS problem. A weaker linear relaxation leads to a looser bound on the optimal solution, potentially increasing the number of branch-and-bound iterations required to converge to optimality. This effect makes the problem more challenging for solvers, as they must explore a larger portion of the search space. Similarly, introducing symmetries increases the number of equivalent solutions, forcing the solver to handle redundant search paths, which can further complicate optimization.

All methods presented in this appendix have been tested with our QPU friendly instances. While they did not show higher effectiveness than the degree centrality method discussed in Section \ref{subsec:Impact of weight}, certain schemes may still be beneficial for other graph structures not considered in this work. Their inclusion remains valuable for completeness.

We present the weighting schemes of each method in what follows.

\begin{enumerate}
    \item \textbf{Matching-based weighting}
    In this method:
    \begin{itemize}
        \item Nodes included in the matching receive a weight of 1.
        \item Nodes excluded from the matching are weighted based on their degree:
        \begin{equation}
        w(v) = 0.1 \times (\text{deg}(v) + 1),
        \end{equation}
        where $\text{deg}(v)$ denotes the degree of node $v$.
    \end{itemize}
    The objective here is to introduce symmetry by ensuring equivalent weight assignment within edges in the matching. This approach may increase the search space, thereby complicating the search for an optimal solution.

    \item \textbf{2-distance matching-based weighting}
    The maximum 2-distance matching problem \cite{balakrishnan2004distance} extends the classical matching problem by enforcing a stricter adjacency constraint. Unlike classical matching, where each node belongs to at most one matched edge, 2-distance matching ensures that matched edges are not adjacent. For certain classes of graphs, we believe that this additional constraint should increase the number of equivalent solutions compared to the simple matching, thereby expanding the search space.

    Solving this problem is NP-complete. The ILP formulation is:
    \begin{align*}
    \text{Maximize} & \quad \sum_{(i,j) \in E} x_{ij} \\
    \text{Subject to} & \quad \sum_{j \in V} x_{ij} \leq 1 \quad \forall i \in V, \\
                      & \quad \sum_{i \in V} x_{ij} \leq 1 \quad \forall j \in V, \\
                      & \quad x_{ij} + x_{ik} \leq 1 \quad \forall (i,j), (i,k) \in E, \text{ s.t. } (j,k) \in E, \\
                      & \quad x_{ij} \in \{0, 1\} \quad \forall (i,j) \in E.
    \end{align*}

    \item \textbf{Degree-based weighting}
    In this method:
    \begin{itemize}
        \item Nodes with the minimum degree receive a weight of 0.1.
        \item Other nodes receive:
        \begin{equation}
        w(v) = 1000 \times (\text{deg}(v) + 1).
        \end{equation}
    \end{itemize}

\item \textbf{Annihilation-based weighting}  
The annihilation number \cite{levit2018annihilation, larson2011graphs} is the largest integer \( k \) such that:  
\begin{equation}
\sum_{i=1}^{k} \deg(v_i) \leq |E|,
\end{equation}  
where \( \deg(v_i) \) denotes the degree of node \( v_i \) (i.e., the number of edges incident to \( v_i \)), and \( |E| \) is the total number of edges in the graph.

In this weighting scheme, nodes belonging to the annihilation set receive a high weight of $1000$, while all other nodes are assigned a small weight of $0.1$.  By assigning high weights to the annihilation set, we aim to negatively impact the quality of the linear relaxation, potentially weakening the bond and increasing the difficulty of solving the problem.

\item \textbf{Maximum clique-based weighting}  
The maximum clique problem \cite{singh2015survey} seeks to identify the largest complete subgraph, where every pair of nodes is connected by an edge. Its ILP formulation is:  
\begin{align*}
\text{Maximize} & \quad \sum_{i \in V} x_i \\
\text{Subject to} & \quad x_i + x_j \leq 1 \quad \forall (i,j) \notin E, \\
                  & \quad x_i \in \{0, 1\} \quad \forall i \in V.
\end{align*}  

In this method, nodes within the maximum clique receive a weight of 1, while all other nodes are assigned a weight of 1000. By assigning identical weights to all nodes in the maximum clique, we can introduce a high degree of symmetry in the problem. Since all clique nodes become equivalent in terms of weight, multiple optimal solutions exist, significantly increasing the number of symmetric solutions that the solver must consider.

    \item \textbf{Graph centrality-based weighting}
    \begin{itemize}
        \item \textbf{Closeness centrality} The closeness centrality of a node $v$ quantifies how close it is to all other nodes in the graph, based on shortest path distances. It is defined as:
        \begin{equation}
C_C(v) = \frac{n-1}{\sum_{u \in V} d(v, u)},
\end{equation}
where:
\begin{itemize}
\item $d(v, u)$ is the shortest path distance between nodes $v$ and $u$.
\item $n$ is the total number of nodes in the network.
\end{itemize}
The weight assignment based on closeness centrality is given by:
\begin{equation}
w(v) = 1 + 999 \times \left(\frac{C_C(v)}{\max\limits_{u \in V} C_C(u)}\right), \quad \forall v \in V.
\end{equation}
Normalizing $C_C(v)$ in this manner ensures that weights remain properly scaled and aligned with the relative centrality of each node. 

\item \textbf{Betweenness centrality}  
The betweenness centrality of a node $v$ measures the extent to which it lies on the shortest paths between other nodes. A node with high betweenness centrality frequently acts as a bridge, meaning that many shortest paths pass through it, influencing the overall connectivity and flow of information in the graph. It is defined as:  
\begin{equation}
C_B(v) = \sum_{s \ne v \ne t} \frac{\sigma_{st}(v)}{\sigma_{st}},
\end{equation}  
where:
\begin{itemize}
    \item $\sigma_{st}$ is the total number of shortest paths from node $s$ to node $t$.
    \item $\sigma_{st}(v)$ is the number of those shortest paths that pass through node $v$.
\end{itemize}
The weight assignment based on betweenness centrality is:
\begin{equation}
w(v) = 0.1 + 999 \times \left(\frac{C_B(v) - \min\limits_{u \in V} C_B(u)}{\max\limits_{u \in V} C_B(u) - \min\limits_{u \in V} C_B(u)}\right), \quad \forall v \in V.
\end{equation}

    \end{itemize}
Graph centrality-based weighting methods exploit topological properties to increase the complexity of the MWIS problem, with the objective of weakening the linear relaxation and making it less effective in providing tight upper bounds.
\end{enumerate}

\end{document}